\documentclass{aa}  
\usepackage{graphicx}
\usepackage{amsmath}
\usepackage{amssymb}
\usepackage{natbib}
\usepackage{booktabs}
\usepackage{multirow}
\usepackage{pdflscape}
\usepackage{float}
\usepackage{subfig}
\usepackage{color}
\usepackage{xcolor}
\usepackage[font=small]{caption}
\usepackage{bm}
\usepackage[normalem]{ulem}

\begin{document}

\offprints{claire.guepin@iap.fr}

\title{Proton acceleration in pulsar magnetospheres}
\authorrunning{C. Gu\'epin, B. Cerutti and K. Kotera}
\date{\today}
\author{Claire Gu\'epin\inst{1,2,3}, Beno\^it Cerutti\inst{4} and Kumiko Kotera\inst{1}}

\institute{Sorbonne Universit\'e, CNRS, UMR 7095, Institut d’Astrophysique de Paris, 98 bis bd Arago, 75014 Paris, France
\and Department of Astronomy, University of Maryland, College Park, MD  20742, USA
\and Joint Space-Science Institute, University of Maryland, College Park, MD  20742, USA
\and Univ. Grenoble Alpes, CNRS, IPAG, 38000 Grenoble, France}

\abstract{Pulsars have been identified as good candidates for the acceleration of cosmic rays, up to ultra-high energies. However, a precise description of the acceleration processes at play is still to be established. Using 2D particle-in-cell simulations, we study proton acceleration in axisymmetric pulsar magnetospheres.
Protons and electrons are extracted from the neutron star surface by the strong electric field induced by the rotation of the star, and electrons and positrons are produced in the magnetosphere through pair production process. As pair production has a crucial impact on electromagnetic fields, on gaps and thus on particle acceleration, we study its influence on the maximum energy and luminosity of protons escaping the magnetosphere. 
Protons are accelerated and escape in all our simulations. However, the acceleration sites are different for the protons and the pairs. As shown in previous studies, pairs are accelerated to their highest energies at the Y-point and in the equatorial current sheet, where magnetic reconnection plays and important role. In contrast, protons gain most of their kinetic energy below the light-cylinder radius within the separatrix current layers, but they are not confined within the equatorial current sheet. Their maximum Lorentz factors can reach $15\%$ to $75\%$ of the maximum Lorentz factor obtained by acceleration through the full vacuum potential drop from pole to equator, and increase with decreasing pair production. Their luminosity can reach $0.2\%$ to $4\%$ of the theoretical spin down luminosity of an aligned pulsar, and the minimum luminosity is obtained at the transition between the force-free and electrosphere regimes. These estimates support that millisecond pulsars could accelerate cosmic rays up to PeV energies and that new born millisecond pulsars could accelerate cosmic rays up to ultra-high energies.}

\keywords{pulsars: general -- acceleration of particles -- methods: numerical}

\maketitle

\section{Introduction}

Pulsars are rapidly rotating and highly magnetized neutron stars, that have been detected across the entire electromagnetic spectrum, from radio to gamma rays (see e.g. \citealp{Abdo13} for a Fermi-LAT catalog of gamma-ray pulsars and \citealp{Aliu08, VERITAS11, Aleksic12} for the detection of the Crab pulsar above $100\,{\rm GeV}$). Their high-energy emissions have been associated with the radiation of accelerated leptons (e.g \citealp{Arons83, Cheng86, Romani96, Muslimov03} for acceleration by unscreened electric fields close to the neutron star surface).

A description is emerging of the structure of the magnetosphere from first principles accounting for the feedback of particles on the electromagnetic fields \citep[e.g.][]{Philippov14, Chen14, Philippov15, Cerutti15, Belyaev15, Brambilla18}. The various particle interactions occurring in the magnetosphere and in particular the production of pairs \citep[e.g.][]{Daugherty82, Gurevich85, Zhang00, Medin10, Timokhin10, Timokhin13} are processes that remain to be fully understood and self-consistently implemented into large-scale systems. This could have a critical influence on our understanding of pair multiplicities and high-energy emissions of pulsars, and might help to improve the models for energy dissipation and spin down. Another fundamental question is related to the nature of the wind around pulsars. The location of the energy dissipation where the Poynting flux is dissipated into particle kinetic energy is still to be clearly identified \citep[e.g.][]{Coroniti90, Kirk03, Komissarov13, Porth13, Cerutti17b}. Finally, the mechanisms for cosmic-ray acceleration in pulsars, studied for instance in \citet{Venkatesan97}, \citet{Blasi00}, \citet{Arons03}, \citet{Fang12,Fang13}, \citet{Lemoine15} and \citet{Kotera15}, should be modelled from first principles in order to more precisely infer  the contribution of these sources to the observed cosmic-ray flux. Many source categories have already been considered as potential cosmic-ray accelerators, such as active galactic nuclei, gamma-ray bursts, tidal disruption events, and pulsars. Despite the advent of time-domain and multimessenger astronomy, together with significant progress in source emission modelling, it is still difficult to discriminate between the various source scenarios.

In this context, a kinetic approach is required. Interestingly, most of the previous studies focused on magnetospheres filled with a plasma of electrons and positrons, without ion injection, and the injection of ions has only recently been considered \citep{Chen14,Philippov18}. It is therefore timely to study the fate of protons in pulsar magnetospheres. \cite{Chen14} study an axisymmetric pulsar magnetosphere, where ions and electrons are injected from the neutron star surface and pairs can be produced. These latter authors notice the different trajectories of pairs and ions as well as the acceleration and escape of ions in different configurations. \cite{Philippov18} consider a similar setup with an oblique rotator. In their simulations, ions have the same mass as positrons but they do not suffer radiative energy losses. They notice the acceleration of ions in the current sheet, mostly at the Y-point.

In the present study, we focus on the two following fundamental questions: (i) what is the maximum energy achievable for the ions, especially for the ones escaping the magnetosphere, and (ii) to what level can they contribute to the observed high- and ultra-high-energy cosmic-ray fluxes? To this end, we perform particle-in-cell (PIC) simulations of the aligned pulsar magnetosphere.  With this series of numerical experiments, we aim to explore the transition between a charge-separated magnetosphere, or `electrosphere' in the following, and a force-free magnetosphere, by changing the yield of pair production, and assessing its impact on particle acceleration and escape. We describe the theoretical and numerical setup in Section~\ref{Sec:setup}. The structure of the magnetosphere is described in Section~\ref{Sec:magnetosphere}. The questions of proton acceleration in the simulations and how our results scale up to realistic pulsar parameters are addressed in Section~\ref{Sec:acceleration}, followed by a discussion and our conclusions in Section~\ref{Sec:conclusion}.

\section{Simulating a pulsar magnetosphere}\label{Sec:setup}

We make use of the PIC code {\tt ZELTRON} \citep{Cerutti13} in its 2D axisymmetric version, with a non-uniform spherical grid, which is well suited to the study of aligned rotators.

\subsection{Electromagnetic fields}

In the following, $r$, $\theta$, and $\phi$ are the usual spherical coordinates. The initial setup of our simulations is a perfectly conducting neutron star in a vacuum, with a magnetic dipole anchored at its surface
\begin{align}
B_r(r,\theta) &= B_\star R_\star^3 \cos\theta / r^3 \, , \\
B_\theta(r,\theta) &= B_\star R_\star^3 \sin\theta / 2 r^3 \, , \\
B_\phi (r,\theta) &= 0 \, ,
\end{align}
where $R_\star$ is the radius of the neutron star, $\theta$ is the angle from the rotation axis, and $B_\star$ is the polar magnetic field. For a perfect conductor rotating at angular velocity $\Omega$, $\bm{E'}=\bm{E}+(\bm{\Omega}\times \bm{r})\times\bm{B}/c =0$ in the co-rotating frame, where $E$ and $B$ are the electric and magnetic fields in the observer frame, and $\bm{\Omega}$ is along the rotation axis. This allows us to estimate the electric field inside the star $(E_r^{\rm int},E_\theta^{\rm int},E_\phi^{\rm int}) = (r\sin\theta/R_{\rm LC}) (B_\theta, -B_r, 0)$, where $R_{\rm LC} = c / \Omega$ is the light cylinder radius defined as the distance at which the corotating speed reaches the speed of light. At $t=0$, the rotation of the neutron star is forced by imposing at its surface the poloidal electric field induced by the rotation of a perfect conductor. The radial electric field can be discontinuous for a non-zero surface charge density. The outer boundary condition is defined by an absorbing layer in order to mimic an open boundary with no information coming back inwards \citep{Birdsall91, Cerutti15}. Apart from these boundary conditions, there are no constraints on the external electric field, which evolves self-consistently during the simulation.

\subsection{Particle extraction}

We note that at the surface of the star, the electric field is of the order $E_\star \sim B_\star R_\star / R_{\rm LC} \sim 10^8 \,{\rm statV \, cm}^{-1}$ for a millisecond pulsar with $B_\star=10^9\,{\rm G}$. Due to this high electric field, charged particles can be extracted from the neutron star surface. In our work, we neglect the molecular or gravitational attraction (\citealp{Petri16}, see however \citealp{Ruderman75}). We consider three particle species: electrons, positrons, and protons. Electrons and protons are extracted from the surface and positrons are created through a pair production process. In order to avoid overinjection, particles can be extracted when the local charge density does not exceed the local Goldreich-Julian (\citealt{Goldreich69}, GJ) charge density $\rho_{\rm GJ}$. At the neutron star surface, for a dipole magnetic field and a rotation around the vertical axis, the GJ charge density is
\begin{equation}
\rho_{\rm GJ} = \frac{ -B_\star (3\cos^2\theta-1) }{  4\pi R_{\rm LC} \left[1- (R_\star \sin \theta / R_{\rm LC})^2 \right]} \,.
\end{equation}
The denominator adds a relativistic correction due to the modification of the magnetic field structure by currents, and is as small as $R_\star / R_{\rm LC} \simeq 0.2$ for a millisecond pulsar. Therefore, for electrons, the GJ number density at the surface of the neutron star reads $n_{\rm GJ} = B_\star (3\cos^2\theta-1)/4\pi R_{\rm LC} e$.

\subsection{Energy losses by radiation}

In our simulations, the motion of a particle is governed by the Abraham-Lorentz-Dirac equation
\begin{equation}
\frac{{\rm d} \bm{p}}{{\rm d}t} = q (\bm{E} + \bm{\beta} \times \bm{B}) + \bm{g} \, ,
\end{equation}
where $\bm{p} = \gamma m \bm{v}$ is the particle momentum, $\gamma = 1/\sqrt{1-\beta^2}$ the particle Lorentz factor, $\bm{v} = \bm{\beta} c$ the particle 3-velocity, $m$ the particle mass, and $q$ the particle electric charge. The first right-hand side term is the usual Lorentz force and $\bm{g}$ is the radiation reaction force due to the radiation of accelerated particles given by the Landau-Lifshitz formula in the framework of classical electrodynamics \citep{LandauLifshitz75}:
\begin{eqnarray}
\bm{g} &=& \frac{2}{3} \frac{q^4}{m^2 c^4} \left[ (\bm{E}+\bm{\beta}\times\bm{B})\times\bm{B} + (\bm{\beta}\cdot\bm{E})\bm{E} \right] \nonumber \\
&&-\frac{2}{3} \frac{q^4 \gamma^2}{m^2 c^4}  \left[ (\bm{E}+\bm{\beta}\times\bm{B})^2 - (\bm{\beta}\cdot\bm{E})^2 \right] \bm{\beta} \,,
\end{eqnarray}
where the terms containing the time derivative of the fields are neglected \citep{Tamburini10}. This simplified view suffices for the current exploration study, as it accounts for synchrotron, synchro-curvature, and curvature regimes; but we note that detailed models that have recently been developed can lead to deviations from the standard curvature and synchrotron radiation spectra in the strong field regime \citep[e.g.][]{Voisin17}.

\subsection{Pair production}\label{section:pairs}

The configuration of the magnetosphere, especially the plasma density and the existence of gaps, relies primarily on the production of electron and positron pairs. The pairs are also thought to contribute to the high-energy radiation of Pulsar Wind Nebulae (PWNe) through synchrotron and inverse-Compton radiation. A precise understanding of the pair production process is therefore critical for the modelling of pulsars. However, the amount of pair production in pulsar magnetospheres is poorly constrained. The pairs are thought to be mainly produced in the polar cap regions \citep{Ruderman75} by the conversion of high-energy gamma rays into pairs in strong magnetic fields (i.e. $B \gtrsim 10^{11}\,{\rm G}$) and the subsequent development of a pair cascade. In the classical model, gamma rays are initially produced through curvature radiation. In the outer gaps, the interaction of gamma-ray photons with X-ray photons from the neutron star surface could also make a significant contribution to the production of pairs in the pulsar magnetospheres \citep{Cheng00}. The pair multiplicity $\kappa = (n^++n^-)/2n_{\rm GJ}$, which describes the number of electron and positron pairs produced by each primary particle, is a poorly constrained parameter that could range between $1$ and $10^7$. From observations and PWNe emission models, the multiplicity has been estimated to be about $10^5-10^7$ for the Crab PWN and $10^5$ for the Vela PWN \citep[e.g.][]{deJager07,Bucciantini11}. However, recent theoretical predictions limit the pair multiplicity to about a few hundreds of thousands \citep{Timokhin18} achieved for magnetic fields of $4\times 10^{12}\lesssim B \lesssim 10^{13}\,{\rm G}$ and hot neutron star surfaces $T\gtrsim 10^6\,{\rm K}$, which questions the existing models of PWNe emissions requiring very high pair multiplicities.

Electron--positron pair plasma generation is the subject of active research \citep{Timokhin13,Chen14}. In this study, we do not aim to model the full pair cascades; a simplified treatment is adopted, as described in \cite{Philippov15}. Pairs are directly produced at the location of the parent lepton if its Lorentz factor exceeds the threshold $\gamma > \gamma_{\rm min, pp}$ and the produced pairs have a Lorentz factor $\gamma_f \sim f_{\gamma} \gamma_i$, which is a fraction $f_{\gamma}=0.1$ of the Lorentz factor of the parent particle $\gamma_i$. This fraction is chosen for numerical reasons, namely to conserve a reasonable separation of scales. The threshold $\gamma_{\rm min, pp} = f_{\rm pp} \gamma_{0, \rm e} $ is a fraction $f_{\rm pp} $ of the maximum Lorentz factor of pairs $\gamma_{0, \rm e} = {e \Phi_0}/{m_e c^2}$ obtained by the acceleration of a particle through the full vacuum potential drop from pole to equator:
\begin{eqnarray}\label{Eq:DDP_tot}
\Phi_0 = -\int_0^{\pi/2} {\rm d}\theta \, R_\star E_\theta(R_\star) = \frac{R_\star^2 B_\star}{2R_{\rm LC}} \, .
\end{eqnarray}
For simplicity, the threshold is constant throughout the magnetosphere, and does not depend on the curvature of the magnetic field lines. This corresponds to a maximization of pair production as all the regions where leptons can be accelerated to sufficiently high energies are active pair-producing regions. Thus, pair production can take place for instance in the equatorial current sheet \citep{Lyubarskii96} and not only in the polar cap regions near the surface of the neutron star. This simplified approach allows us to explore various regimes of the magnetosphere by adjusting the parameter $f_{\rm pp}$, without entering into a detailed modelling of the radiative backgrounds that could significantly contribute to the production of pairs, and would require additional parameters. The connection between the implemented parameter $f_{\rm pp}$ governing the pair production in the entire magnetosphere, and the pair multiplicity $\kappa$ at the polar cap is intricate. A straightforward comparison can be made by computing $\kappa$ at the poles directly in the simulation and comparing the results with theoretical and observed values. Moreover, the amount of pair production in realistic populations of millisecond or newborn pulsars remains to be explored. The modelling of gamma-ray emissions could also help to establish a clearer link between these quantities.

Several additional effects such as photohadronic interactions could impact the particle motion and contribute to energy losses and pair production. For instance, we do not account for Bethe-Heitler processes that could contribute to the production of pairs. Inverse-Compton scattering could also lead to significant energy losses. These effects could be included in a future study.

\subsection{Simulation features}

\begin{table}
\begin{center}
\caption{Simulation parameters.}\label{tab:params}
\resizebox{0.48\textwidth}{!}{
\begin{tabular}{ll}
\toprule  \textbf{Quantity} & \textbf{Estimate} \\
\midrule Neutron star radius &  $R_\star=10^2\,{\rm cm}$  \\
Light cylinder radius &  $R_{\rm LC}=5R_\star$  \\
Polar magnetic field &  $B_\star = 1.1 \times 10^5\,{\rm G}$  \\
Effective magnetic field &  $B_{\rm eff} \simeq 10^9\,{\rm G}$  \\
Proton to electron mass ratio  &  $m_{\rm r} = m_{\rm p}/m_{\rm e} = 18.36$  \\
Polar GJ number density  &  $n_{\rm GJ}^\star = B_\star / 2\pi R_{\rm LC} e$  \\
Polar GJ current density  &  $J_{\rm GJ}^\star = c B_\star / 2\pi R_{\rm LC}$  \\
Smallest cell size  &  $\Delta r^{\star} \simeq 0.13 \,{\rm cm} $  \\
Electron plasma skin depth  &  $d_e^\star \simeq 2 \,{\rm cm} $  \\
Electron gyroradius  &  $r_{{\rm g},e}^\star \simeq 0.015 \,{\rm cm} $  \\
Proton gyroradius  &  $r_{{\rm g},p}^\star \simeq 0.28 \,{\rm cm} $  \\
Polar cap angle  &  $\theta_{\rm pc} \simeq 0.46\,{\rm rad}$  \\
Full potential drop &  $\Phi_0 = R_\star^2 B_\star / 2R_{\rm LC} $  \\
Polar cap potential drop &  $\Phi_{\rm pc} = R_\star^3 B_\star / 2R_{\rm LC}^2 $  \\
Max. proton Lorentz factor &  $\gamma_{0,\rm p}  \simeq 36 $  \\
Max. pair Lorentz factor &  $\gamma_{0,\rm e}  \simeq  667 $\\
Polar cap proton Lorentz factor &  $\gamma_{\rm pc,\rm p} \simeq 7$  \\
Polar cap pair Lorentz factor &  $\gamma_{\rm pc,\rm e} \simeq  133$ \\
Pair production parameter &  $f_{\rm pp}$ \\
\bottomrule
\end{tabular}}
\end{center}
\end{table}

\begin{figure*}[htpb]
\centering
\includegraphics[width=\textwidth]{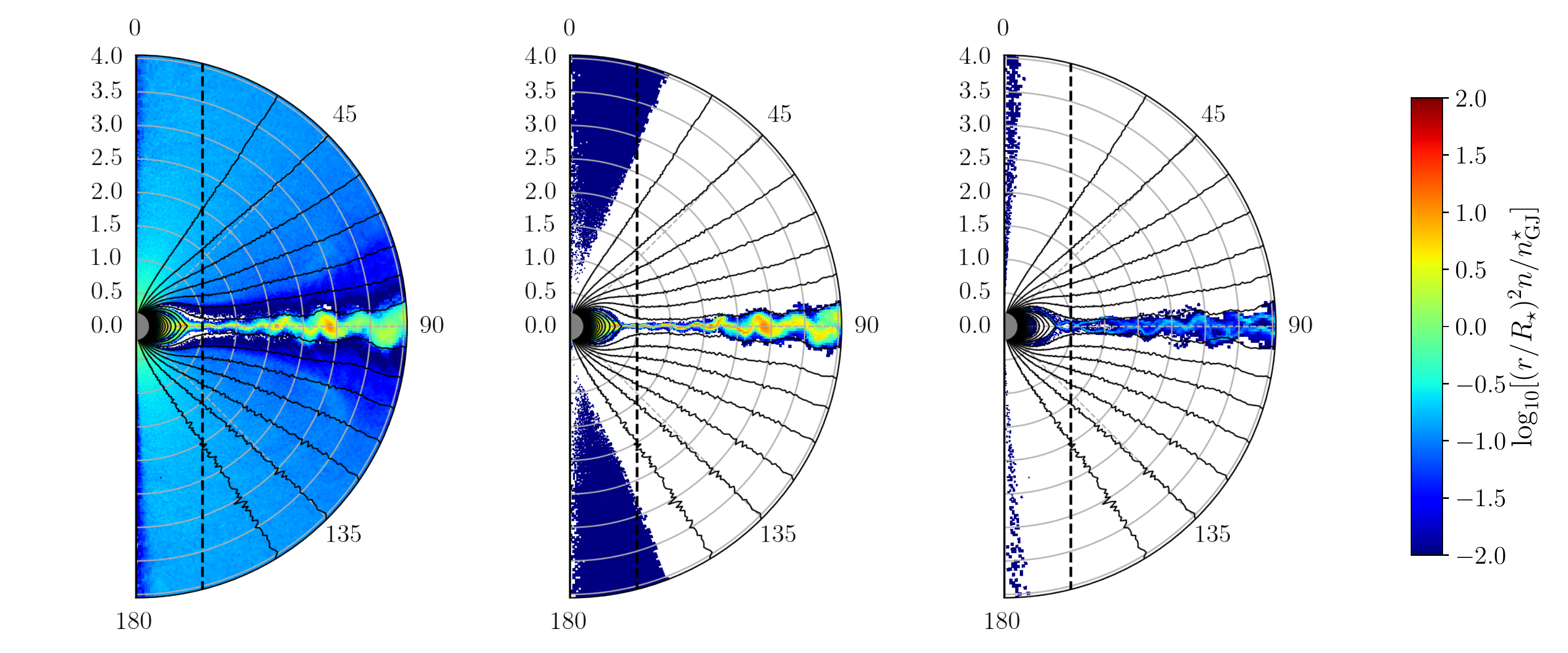}
\includegraphics[width=\textwidth]{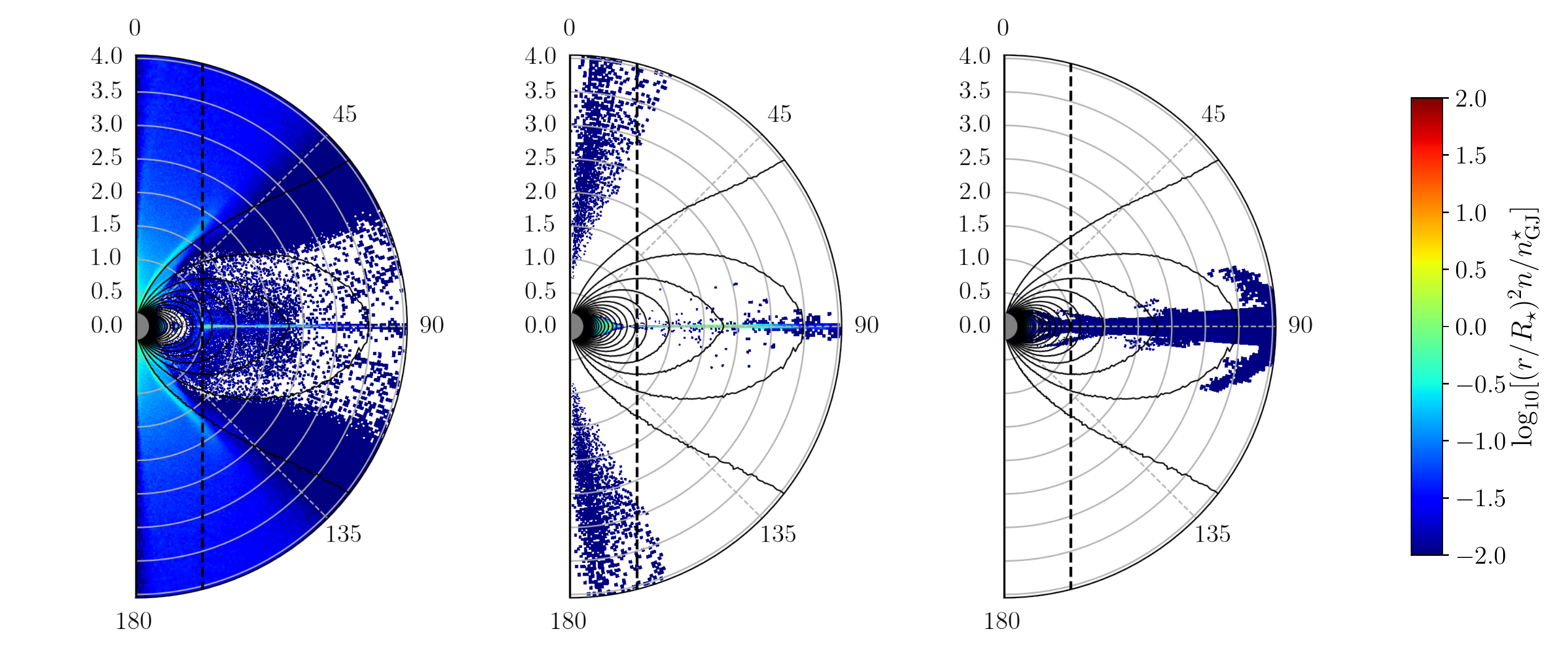}
\includegraphics[width=\textwidth]{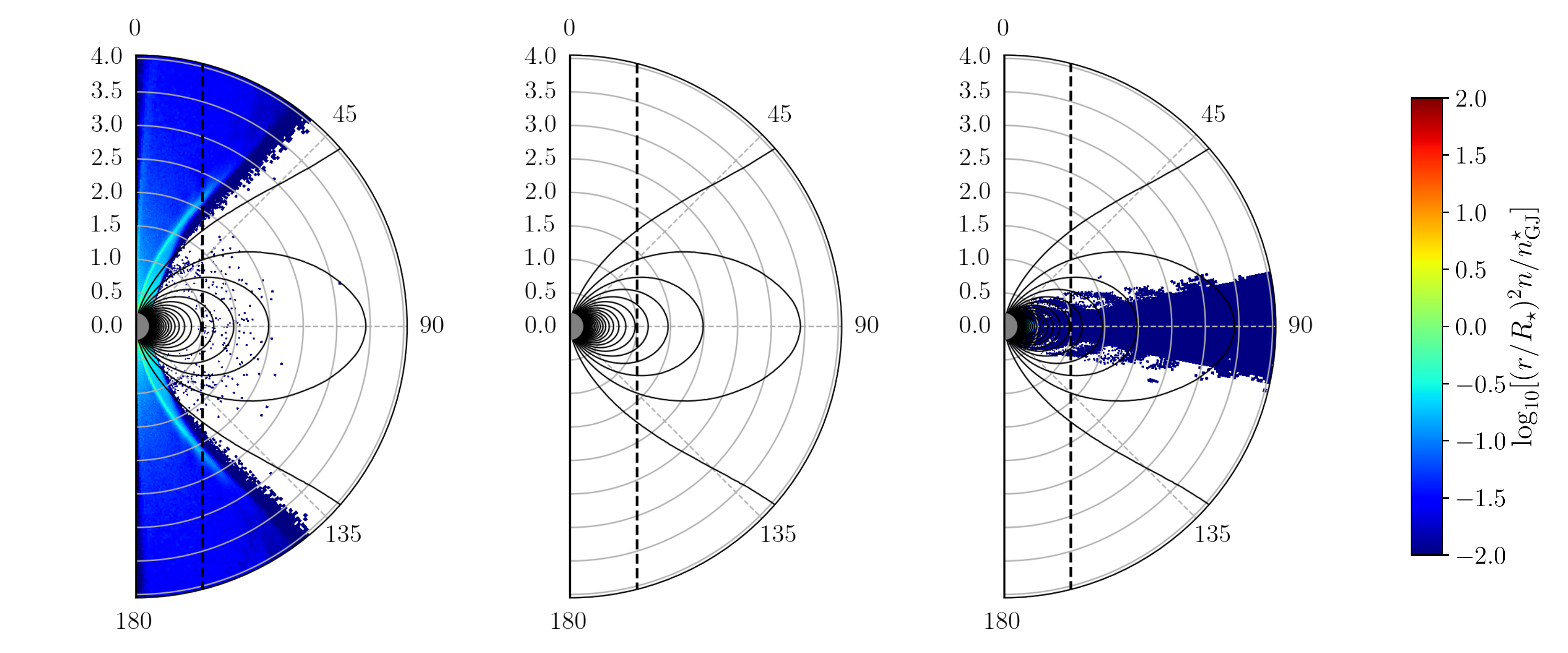}
\caption{Density maps for electrons, positrons, and protons (left to right), for $f_{\rm pp} = 0.01$, $f_{\rm pp} = 0.05$ and $f_{\rm pp} = 0.1$ (top to bottom), as a function of $r/R_{\rm LC}$ and $\theta$, for $t=5P$. The densities are normalised by $ (r/R_\star)^{-2} n_{\rm GJ}^\star$, where $n_{\rm GJ}^\star = B_\star / 2\pi R_{\rm LC} e$ is the polar GJ number density. Solid black lines are the magnetic field lines. The dashed black line indicates the distance from the rotation axis $r \sin\theta = R_{\rm LC}$ and the grey semi-disc represents the neutron star. The densities are in log scale in order to enhance the contrast.}\label{fig:mag_dens}
\end{figure*}

\begin{figure*}[htpb]
\centering
\includegraphics[width=\textwidth]{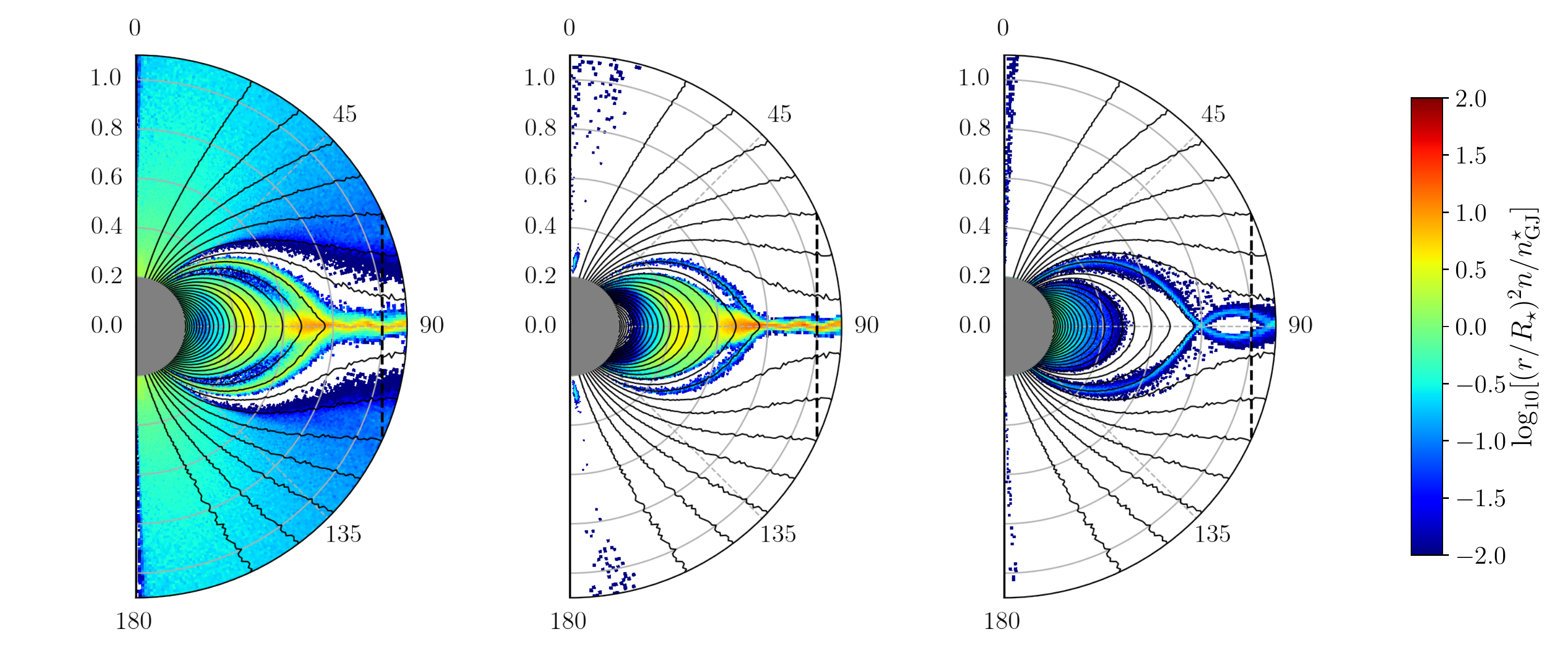}
\includegraphics[width=\textwidth]{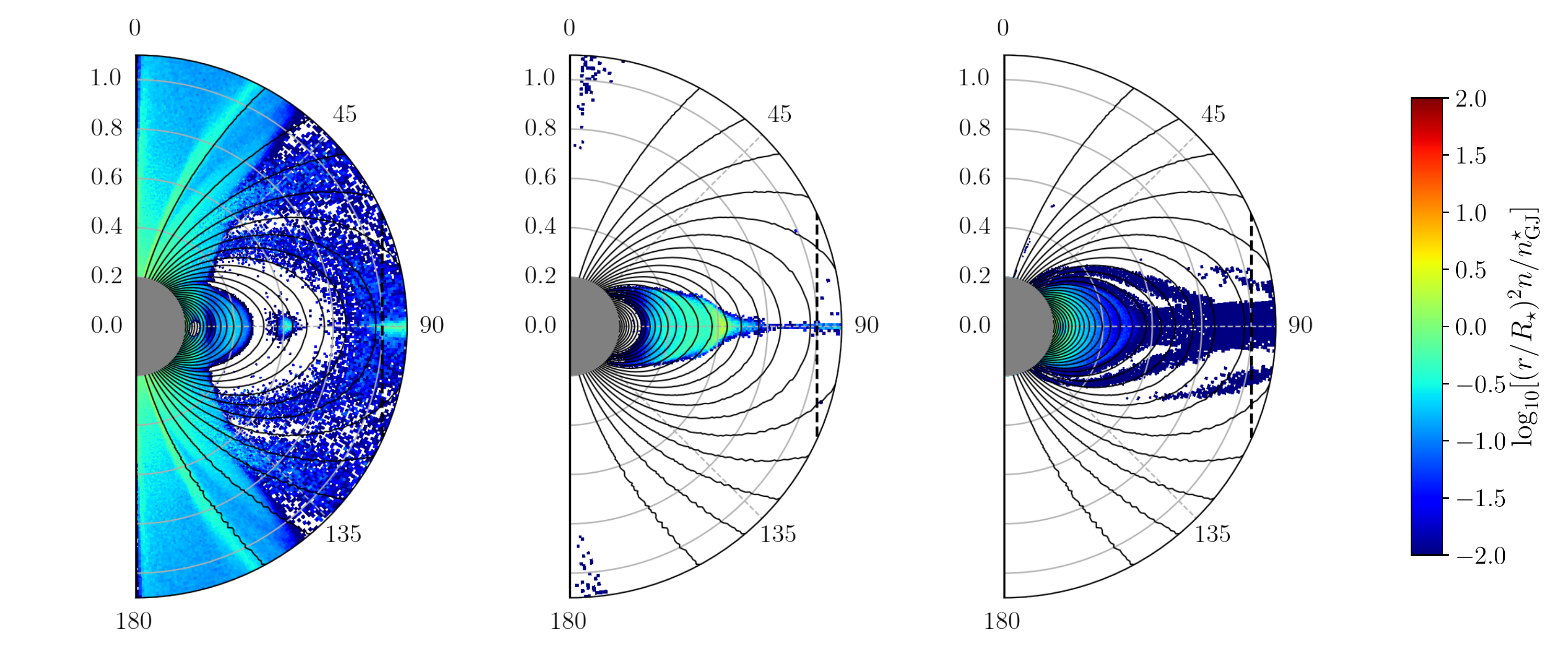}
\includegraphics[width=\textwidth]{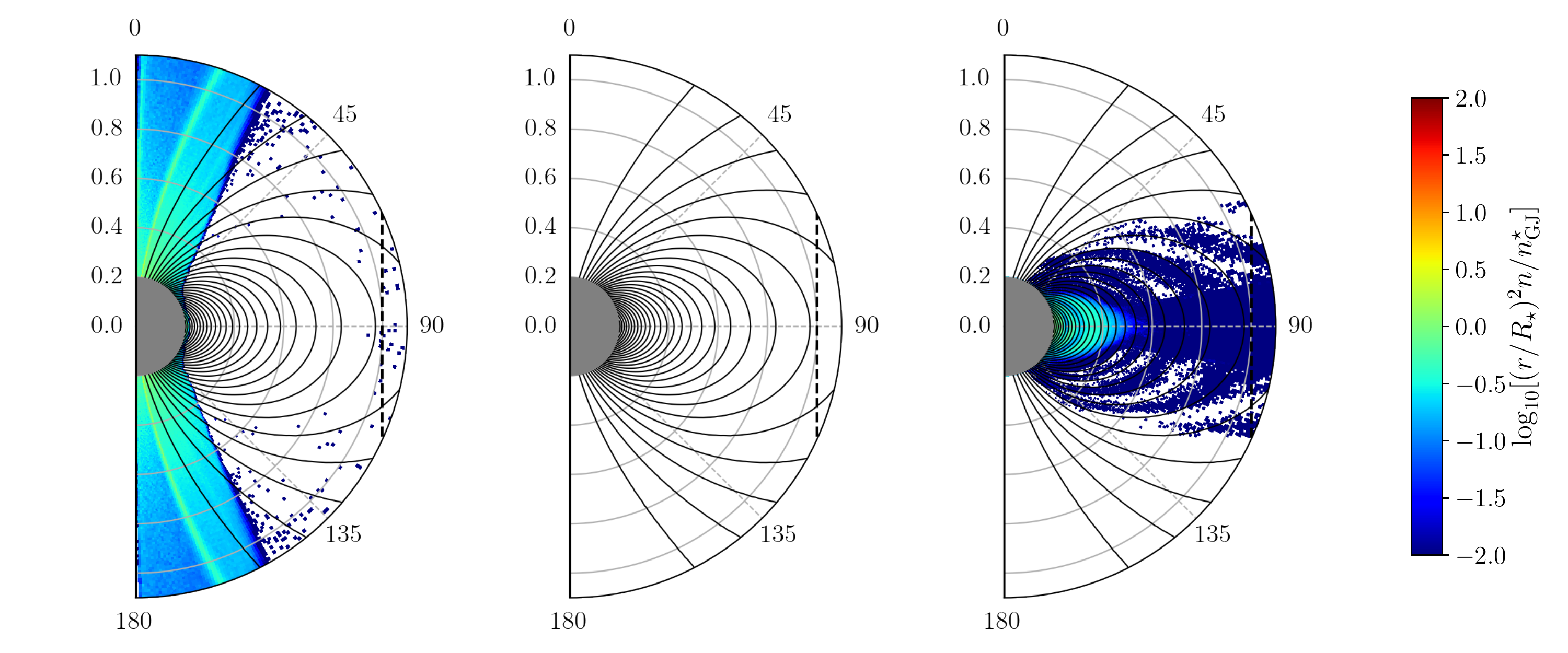}
\caption{Same as figure~\ref{fig:mag_dens} but for the zoomed-in simulation domain $r<R_{\rm LC}$.}\label{fig:mag_dens_zoom}
\end{figure*}

The parameters of our simulations and notations are summarised in Table~\ref{tab:params}. In order to maintain acceptable computation costs, the radius $R_\star$, the magnetic field $B_\star$, and the mass ratio $m_{\rm r} = m_{\rm p}/m_{\rm e}$ are scaled down with respect to realistic values $R_\star \sim 10^6 \,{\rm cm}$, $B_\star \sim 10^8-10^{15}\,{\rm G}$ and $m_{\rm p}/m_{\rm e} \simeq 1836$. In our simulations, $R_\star = 10^2\,{\rm cm}$,  $B_\star = 1.1 \times 10^5\,{\rm G}$ and $m_{\rm r} = 18.36$. However, several typical scales are preserved. For millisecond pulsars, we have $R_{\rm LC}/R_\star = c P / 2\pi R_\star \sim 5 \,P_{-3}\,R_{\star,6}^{-1}$. We adopt this typical value of $R_{\rm LC}/R_\star$ in the simulations. Considering $n_e = n^\star_{\rm GJ}$, the plasma skin depth of electrons at the star surface is $d_e^\star = c / (4\pi n_e e^2/m_e)^{1/2} \simeq2 \,{\rm cm}\,B_9^{-1/2} P_{-3}^{1/2}$ for millisecond pulsars. This value is similar in our simulations, and is resolved by several computational cells.  The gyroradius of electrons $r_{\rm g} = \gamma m c v_\perp / e B$, where $v_\perp$ is the speed perpendicular to the magnetic field direction, is not resolved in our simulations for $ \gamma v_\perp/c \sim 1$, that is, before they are accelerated close to the neutron star surface. This should not have an impact on the simulations, as electrons efficiently lose their perpendicular energy due to strong synchrotron radiation close to the neutron star surface. The radiation reaction force is amplified (within the limits of time resolution) by considering an effective magnetic field $B_{\rm eff} \sim 10^9\,{\rm G}$ in order to reduce the synchrotron cooling time. 

At the beginning of the simulation, the magnetosphere is empty. The system is then strongly perturbed due to the sudden induced electric field, and the subsequent injection of particles and reconfiguration of electromagnetic fields. In order to capture the system behavior when the stationary regime is established, we evolve the system during at least five rotation periods. We check that the simulations have reached a steady state by looking at the temporal evolution of several key quantities, such as the total energy, the radial Poynting flux, and the electric charge of the neutron star. A non-uniform spherical grid of  $2464\times2464$ cells is used; these are logarithmically spaced in $r$ between  $r=R_{\star}$ and $r=5 R_{\rm LC}$, and uniformly in $\theta$ between $\theta=0$ and $\theta=\pi/2$.

\section{Structure of the magnetosphere}\label{Sec:magnetosphere}

\subsection{Plasma density}

\begin{figure}[t]
\centering
\includegraphics[width=0.49\textwidth]{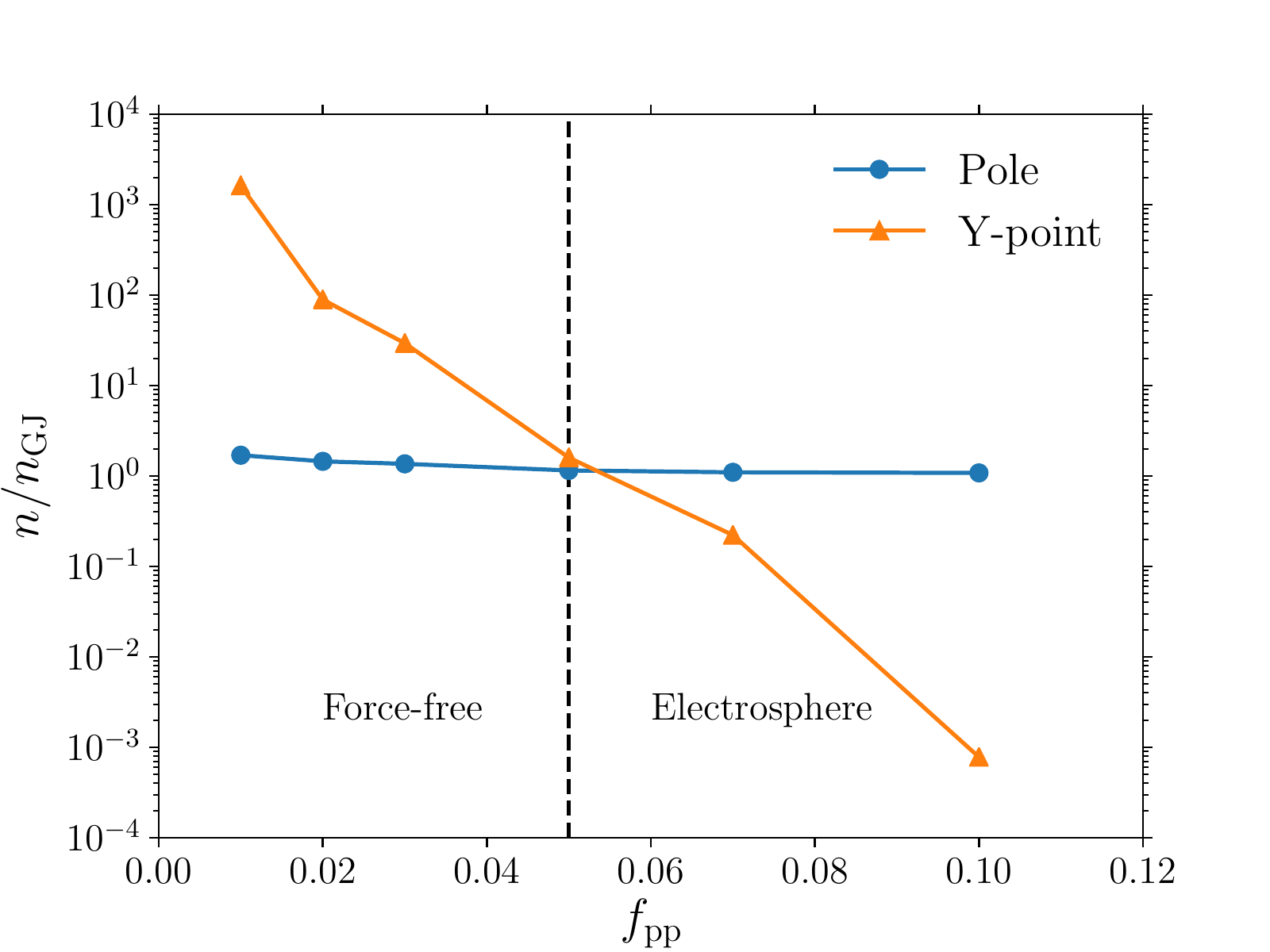}
\caption{Plasma densities $(n^+ + n^-)/ n_{\rm GJ}$ as a function of $f_{\rm pp}$ at the pole and at the Y-point (respectively blue and orange lines).}\label{fig:kappa_comp}
\end{figure}

We study  the impact of changing the amount of pair production on the structure of the magnetosphere. Two extreme regimes of the magnetosphere have been described in the literature: the electrosphere, or disc-dome configuration \citep{Krause85,Smith01,Petri02,Spitkovsky02}, and the force-free configuration \citep{Goldreich69,Contopoulos99,Timokhin06}. In the first configuration, the production of pairs is not sufficient to completely populate the magnetosphere with plasma. Charges are separated in different regions, with a dome of negative charges and a disc of positive charges (if $\Omega\cdot B>0$). In the second configuration, interesting features appear such as the transition between closed field lines and open field lines, the direct volume currents at the poles and the return currents along the last closed field line (the equatorial current sheet). The guideline studies illustrated in Figs.~\ref{fig:mag_dens} and \ref{fig:mag_dens_zoom} mimic the transition between these extreme regimes. These examples are obtained for a high production of pairs ($f_{\rm pp} = 0.01$) which mimics a force-free case and for a low production of pairs ($f_{\rm pp} = 0.05$ and $f_{\rm pp} = 0.1$) tending towards an electrosphere configuration, when a steady state is reached.

Low values of $f_{\rm pp}$ lead to a strong production of pairs and thus allow to study magnetospheres close to the force-free regime. We note that small gaps separate the polar flows and the current sheet. The magnetic field, initially in a dipolar configuration, is strongly affected by the dense plasma outflow. A closed magnetic field line region is maintained at low latitudes below the light-cylinder radius, whereas magnetic field lines open up at high latitudes, which is similar to the configuration in the force-free regime \citep[e.g.][]{Contopoulos99,Timokhin07}. Theoretically, the magnetic field lines have been predicted to present a Y-shape at the point where the last closed field line intersects the equatorial plane (called the Y-null point or Y-point), which seems to be the case in these simulations. We note however that in our simulations, the Y-point is located slightly below the light cylinder radius. We note that in the perfect steady state force-free MHD view, the Y-point is located at the light cylinder. However, in general, it does not have to be so and its location depends on the plasma content and magnetisation, as mentioned for instance in \cite{Belyaev15}. We also observe that the position of the Y-point is highly time-dependent; it has a breathing motion around the light-cylinder radius which reflects the intermittent nature of reconnection. This phenomenon is emphasised by the production of plasma overdensities trapped in magnetic loops, i.e. plasmoids, within the equatorial current sheet (see top panels in figure~\ref{fig:mag_dens}).

Our simulation with $f_{\rm pp}=0.01$ is characterised by number densities of electrons in the polar regions of the order of the polar GJ number density $n_{\rm GJ}^\star = B_\star / 2\pi R_{\rm LC} e$ multiplied by the distance scaling factor $(R_\star / r)^2$. Therefore, only a small number of pairs are produced at the poles in agreement with previous studies \citep{Chen14,Philippov15}. In the equatorial region (the current sheet), the number density of pairs can reach higher values locally, above $10\,n_{\rm GJ}^\star$, which is a signature of strong pair production. Protons are propagating in the equatorial region, with number densities around $1$ to $10\%$ of $(R_\star / r)^2 n_{\rm GJ}^\star $. Below the light cylinder, escaping protons flow along the last closed field line (the separatrices). At larger radii, protons form an equatorial flow, which is not confined inside the current sheet formed by the pairs, because of their larger gyroradii. We note that at the Y-point, the pair and proton skin depths are respectively $d_e/R_\star = \sqrt{\gamma_{\rm pc, e} m_e c^2 / 4\pi n_e e^2}/R_\star \simeq 0.1$ and $d_{\rm p}/R_\star = \sqrt{\gamma_{\rm pc, p} m_{\rm p} c^2 / 4\pi n_{\rm p} e^2}/R_\star \simeq 4$, considering the polar cap Lorentz factors of pairs and protons and the number densities at the Y-point from the simulations. Therefore, protons are sensitive to larger field structures.

\begin{figure}[!t]
\centering
\includegraphics[width=0.49\textwidth]{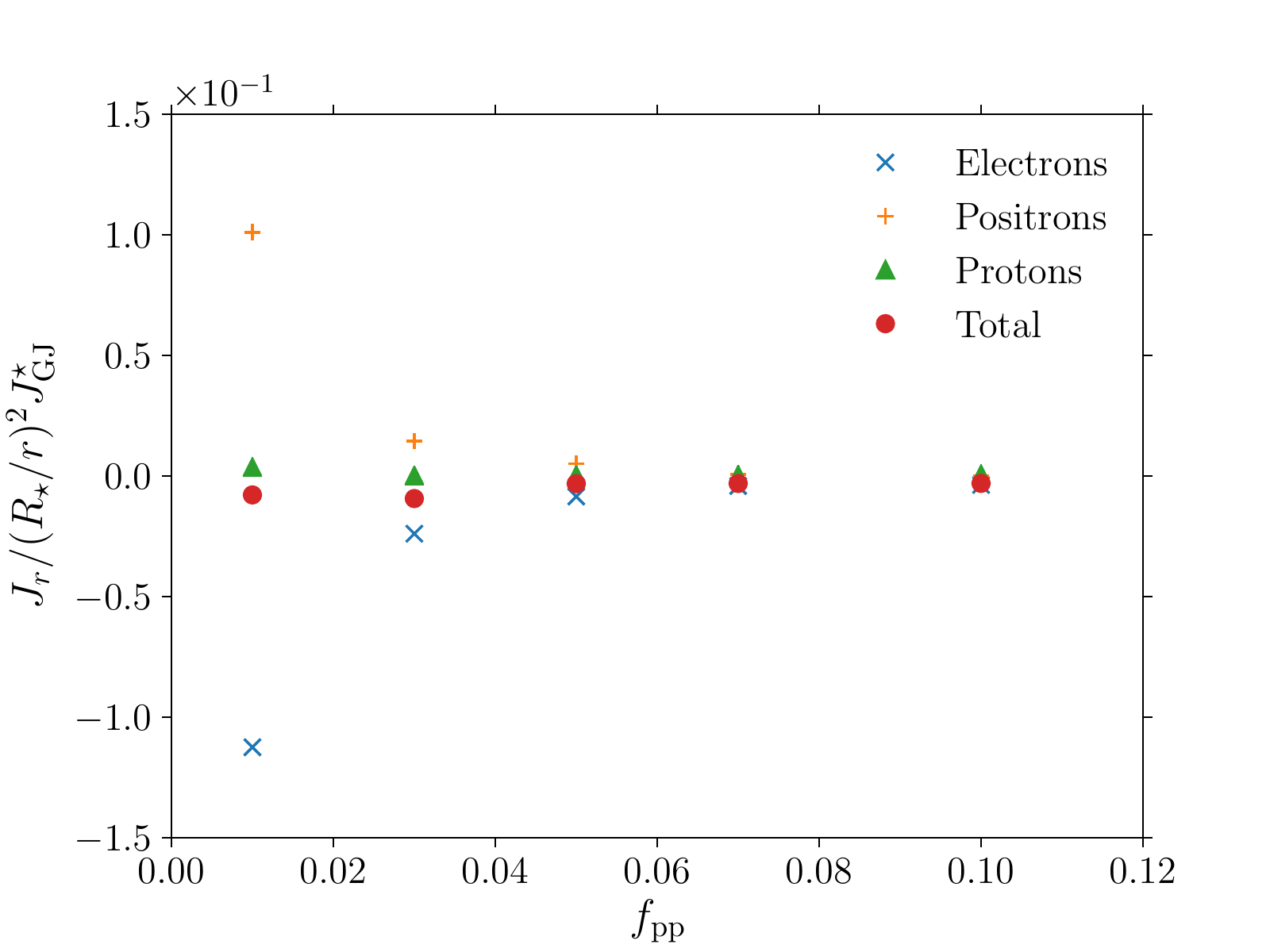}
\caption{Radial current density for $r=2R_{\rm LC}$, averaged over $\theta$ and normalised by the polar GJ current density $(R_\star/r)^2 J_{\rm GJ}^\star$, as a function of $f_{\rm pp}$. We show the total radial current density  (circles) and the contributions of electrons (crosses), positrons (plus signs), and protons (triangles).}\label{fig:current_fpp}
\vspace{0.2cm}
\includegraphics[width=0.49\textwidth]{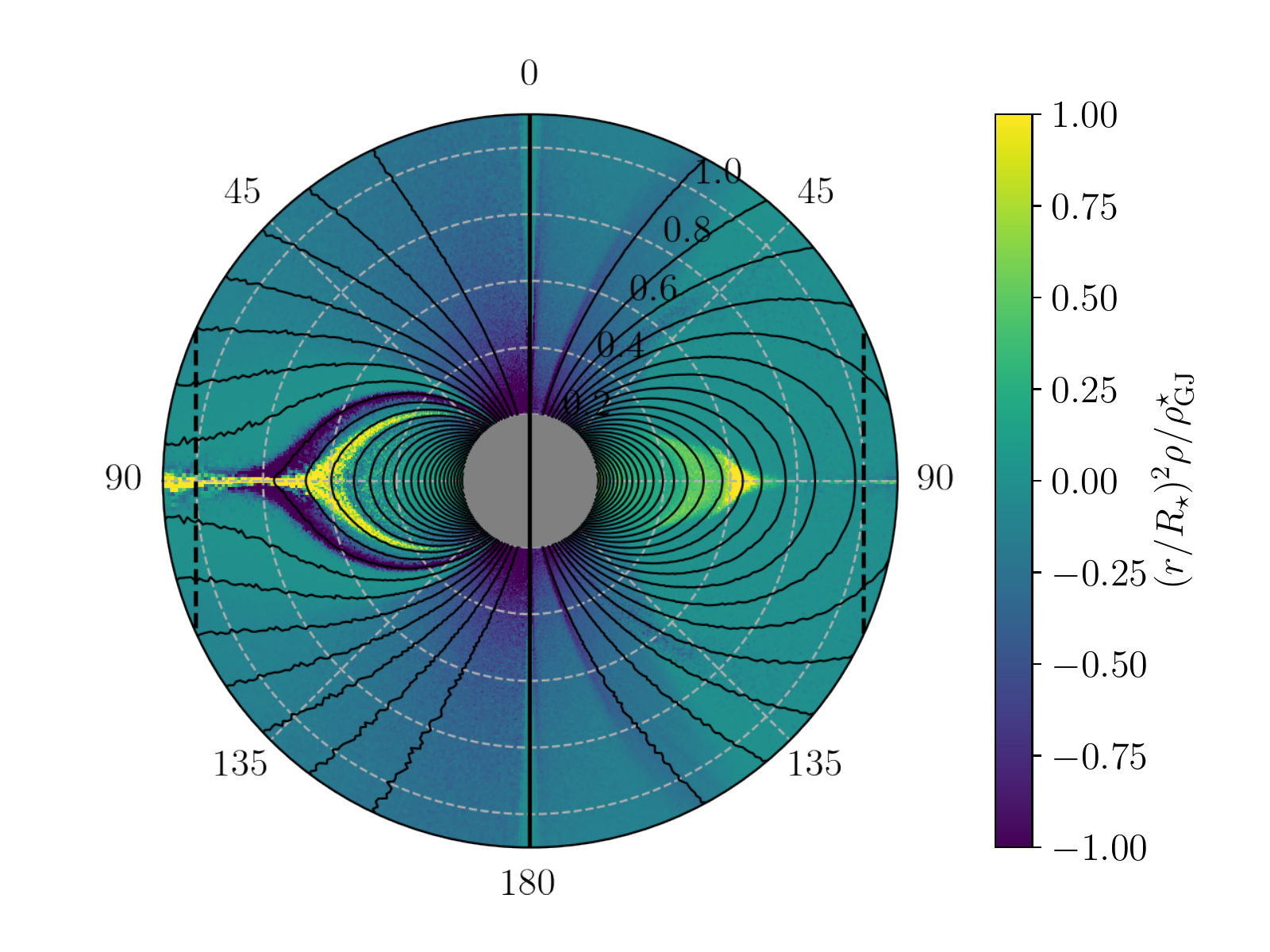}
\includegraphics[width=0.49\textwidth]{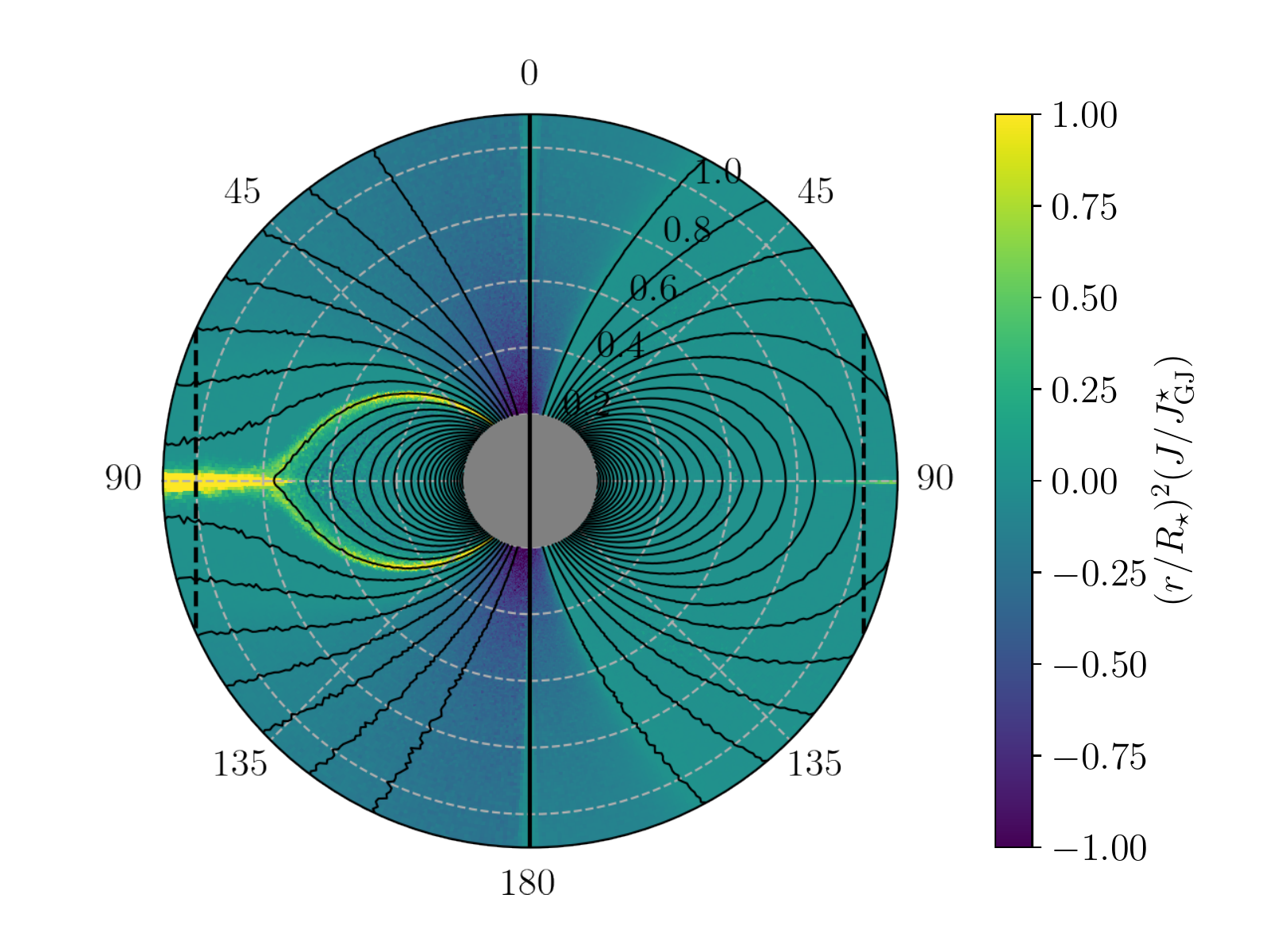}
\caption{Charge density maps (top) and radial current maps (bottom) for a high production of pairs $f_{\rm pp} = 0.01$ (left) and a low production of pairs $f_{\rm pp} = 0.05$ (right), for $r \leq R_{\rm LC}$. $\rho_{\rm GJ}^\star$ and $J_{\rm GJ}^\star$ are the GJ charge and current densities at the poles.}\label{fig:rho_tot}
\vspace{-1cm}
\end{figure}

For high values of $f_{\rm pp}$, the production of pairs is almost absent and allows us to study magnetospheres close to the electrosphere configuration. The simulation with $f_{\rm pp}=0.05$ illustrates the transition between the force-free and electrosphere regimes, and the simulation with $f_{\rm pp}=0.1$ illustrates the electrosphere regime. The magnetic field lines that open up at the beginning of the simulation because of the transitory dense plasma outflow tend to return to a dipolar configuration because of the low plasma density in the stationary regime. High electric fields contribute to maintaining the equatorial flow of protons, but the subsequent current is not sufficiently large to modify the dipolar structure of the magnetic field. The few open field lines are anchored to the star near the poles, where the electrons are extracted. In our simulation with $f_{\rm pp} = 0.1$, there are no positrons in the magnetosphere. Electrons are essentially confined in the polar regions and are characterised by number densities of around $10\%$ of $ (R_\star / r)^3 n_{\rm GJ}^\star$, with higher number densities close to the star surface (as expected $ \sim n_{\rm GJ}^\star$ at the neutron star surface) and in high-latitude elongated regions. In these regions, it appears that electrons are trapped and are going back and forth before escaping or falling back to the star surface. Large gaps of densities separate the bulk of electrons and protons. A high density of protons is confined near the neutron star surface, with $n \sim n_{\rm GJ}^\star$. Low number densities of protons, that is, below $10^{-3}  (R_\star / r)^2 n_{\rm GJ}^\star$, propagate in the equatorial region and along the separation region between the bulk of electrons and the gaps. Due to the structure of the magnetic and electric fields, these protons escape from the disc of proton and swirl around the neutron star, with a large radial velocity.

\subsection{Pair multiplicity at the pole and Y-point}

The plasma densities estimated at the pole and at the expected location of the Y-point are illustrated in figure~\ref{fig:kappa_comp}. The density of pairs divided by the typical local GJ density $n_{\rm GJ} = |\bm{B}\cdot \bm{\Omega}|/2\pi e c $ (where the correction due to the modification of the magnetic field structure by currents is not accounted for) gives a local estimate of the pair multiplicity $\kappa \sim (n^+ + n^-)/ 2 n_{\rm GJ}$, where $n^+$ and $n^-$ are respectively the densities of positrons and electrons. As noted previously, pair production is sub-dominant by several orders at the poles and occurs predominantly in the current sheet, which is consistent with several recent studies \citep{Chen14,Philippov15}. At the pole, $f_{\rm pp} \leq 0.01$ leads to a pair multiplicity of $\kappa \sim 1$, which decreases slightly with increasing $f_{\rm pp}$. At the Y-point, a high production of pairs leads to high multiplicities, for instance $\kappa \sim 10^3$ for $f_{\rm pp} = 0.01$. When the production of pairs decreases, the pair multiplicity drops significantly, below $\kappa = 1$ for $f_{\rm pp} \geq 0.05$.

\subsection{Charge and current densities}

\begin{figure}[!t]
\centering
\includegraphics[width=0.49\textwidth]{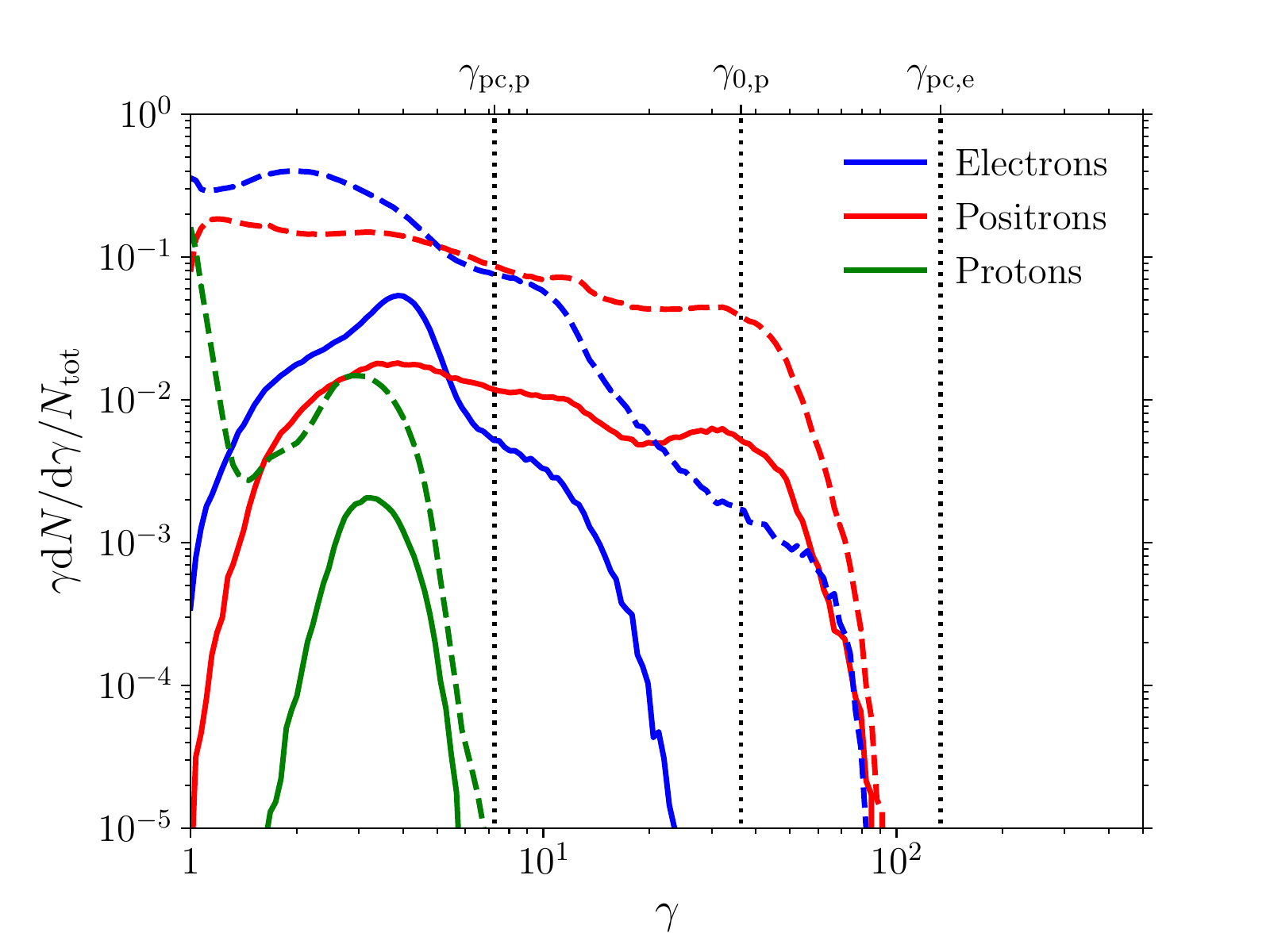}
\includegraphics[width=0.49\textwidth]{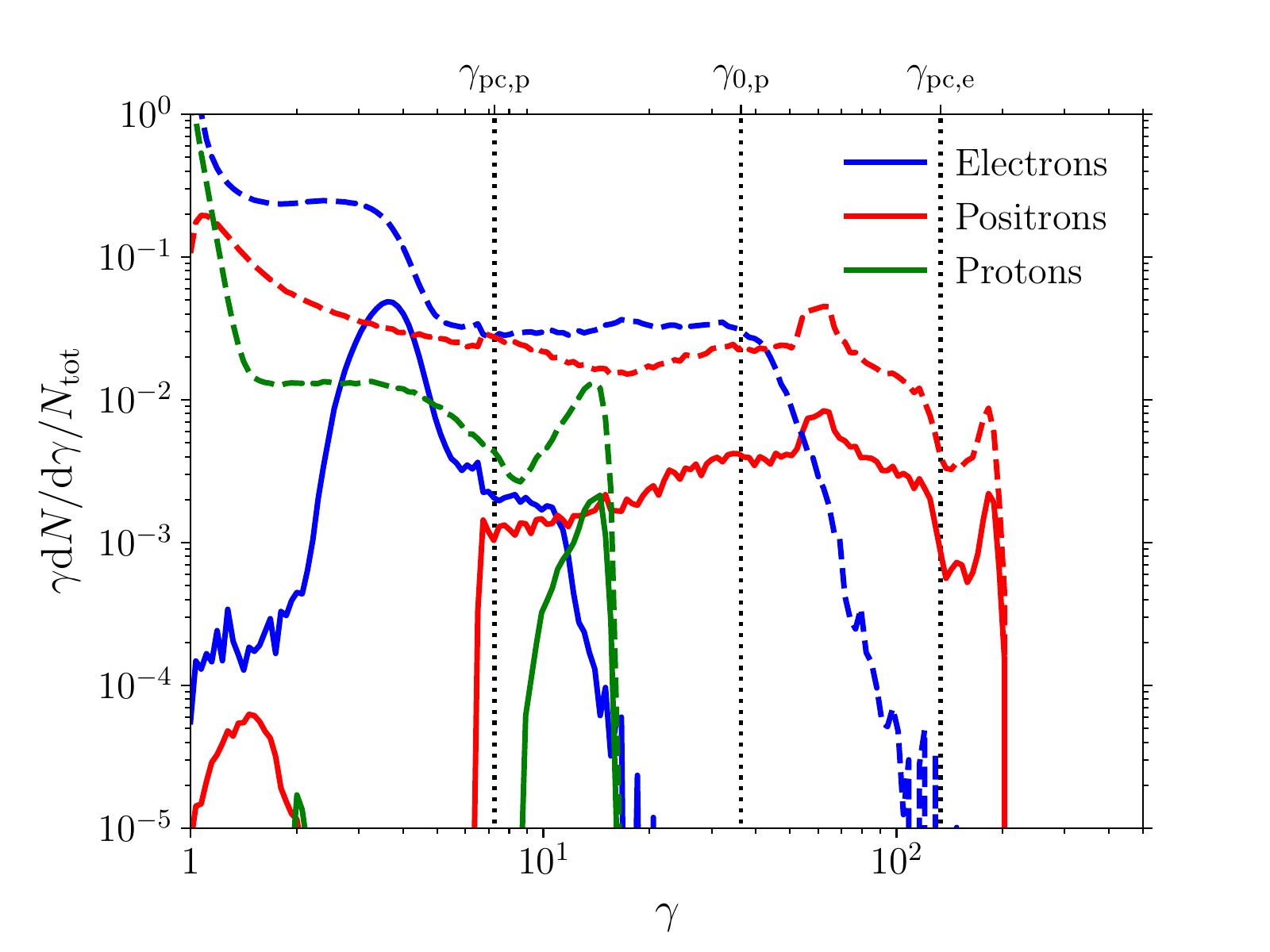}
\includegraphics[width=0.49\textwidth]{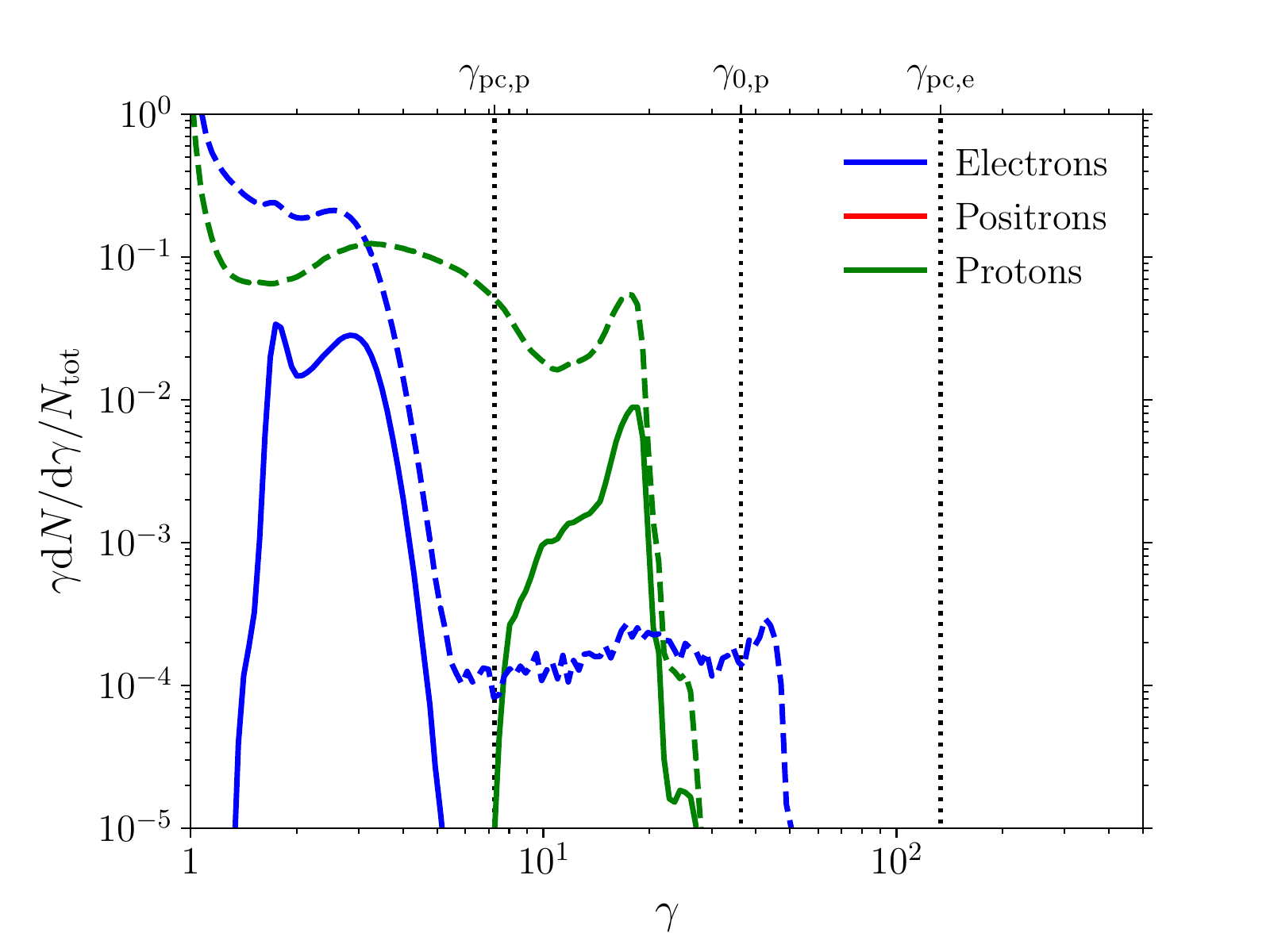}
\caption{Spectra $\gamma {\rm d} N / {\rm d}\gamma$ normalised by the total number of particles in the magnetosphere $N_{\rm tot}$, for electrons (blue), positrons (red), and protons (green), for all particles (dashed lines) and for particles escaping the magnetosphere (solid lines). We compare high, medium, and low production of pairs, $f_{\rm pp}=0.01$, $f_{\rm pp}=0.05,$ and $f_{\rm pp}=0.1$,  respectively (top to bottom). Dotted lines indicate the maximum Lorentz factors $\gamma_{\rm pc,p}$, $\gamma_{\rm 0,p}$ and $\gamma_{\rm pc,e}$.}\label{fig:PIC_SpecOut}
\vspace{-0.1cm}
\end{figure}

The charge densities and radial currents are illustrated in figure~\ref{fig:rho_tot} and have been normalised by the typical GJ charge and current densities at the poles, $\rho_{\rm GJ}^\star = B_\star / 2\pi R_{\rm LC}$ and $J_{\rm GJ}^\star = c B_\star / 2\pi R_{\rm LC} $,  respectively, and multiplied by $(r/R_\star)^2$. They present interesting features, due to the mixing of particle species. For $f_{\rm pp}=0.01$, the poles are dominated by negative charge densities, which carry a negative radial current out of the polar caps. The equator is mostly dominated by positive charge densities, and a positive current density. Just above the last closed field line, a small region is dominated by negative charge densities. The corresponding radial current shows that these negative charges are the main contributors to the return current (they carry a positive radial current), closing on the polar caps. The closed field line region is dominated by positive charge densities, which do not seem to contribute much to the radial current. For $f_{\rm pp}=0.05$ the situation is simpler, with negative charge densities at the poles and positive charge densities at the equator. The charge densities are smaller than for $f_{\rm pp}=0.01$, except in elongated regions around the poles for negative charge densities, and in a disc close to the neutron star surface for positive charge densities. Interestingly, only high latitudes seem to contribute to radial currents, with a small return current directly next to the negative current. Protons and electrons in this region therefore seem to contribute more to the return current than protons in the equatorial region.

Moreover, we illustrate in figure~\ref{fig:current_fpp} the impact of pair production on the radial current density at $r = 2 R_{\rm LC}$ averaged over $\theta$ and normalised by $(R_\star/r)^2 J_{\rm GJ}^\star$. As expected, the total radial current density is close to zero, that is, the star does not charge up.  Electrons and positrons carry respectively most of the negative and positive radial currents, and protons only contribute marginally to the radial current. The radial current densities associated with electrons and positrons decrease for decreasing pair production, with a maximum value of $J_r (r) \sim 0.1 (R_\star/r)^2 J_{\rm GJ}^\star$ at $r = 2 R_{\rm LC}$, for $f_{\rm pp} = 0.01$.

\section{Particle acceleration and energy dissipation}\label{Sec:acceleration}

A magnetized rotating conductor develops a potential difference between the pole and the equator. Particles that experience all or a fraction of the voltage drop can get accelerated through unipolar induction. This is the case for rotating and magnetized neutron stars, that are considered as perfect conductors in our model. As shown in section~\ref{section:pairs}, particles can be accelerated up to $\gamma_0 = {e \Phi_0}/{mc^2}$  where $\Phi_0 = B_\star R_\star^2 / 2 R_{\rm LC}$ (see equation~\ref{Eq:DDP_tot}) if they experience the full vacuum potential drop. A typical fraction of this full vacuum potential drop is given by the potential drop across the polar cap, the surface of the neutron star on which open field lines are anchored. As the typical polar cap angle is $\sin^2 \theta_{\rm pc} \sim R_\star/R_{\rm LC}$, it yields $\Phi_{\rm pc} = { B_\star R_\star^3} /{2R_{\rm LC}^2}$ and $\gamma_{\rm pc} = e \Phi_{\rm pc} / m c^2$. The detail of particle trajectories and structure of the electromagnetic field is important to precisely characterise their acceleration, which is the aim of our simulations. However, these theoretical estimates are useful to better understand and rescale the simulation outputs.

\subsection{Particle spectra}

The spectra of electrons, positrons, and protons are illustrated in Fig.~\ref{fig:PIC_SpecOut} for $f_{\rm pp}=0.01$, $f_{\rm pp}=0.05,$ and $f_{\rm pp}=0.1$. We compare the spectra obtained for all particles in the magnetosphere and for particles escaping the magnetosphere. To compute the spectra of escaping particles, we calculate the total number of particles comprised in the spherical shell between $0.8 \, r_{\rm max}$ and $0.9 \, r_{\rm max}$, such as $u_r>0$. Thus, the normalisations of the total and escaped spectra are only indicative and should not be compared.

Each particle species shows different spectra, as they experience different fates in the magnetosphere. Moreover, the spectra obtained for a high and low pair production show large discrepancies. We note that for $f_{\rm pp}=0.01$, close to the force-free regime, the highest energy electrons and positrons are accelerated to Lorentz factors close to the vacuum polar cap Lorentz factor or pairs $\gamma_{\rm pc, e}$. The electrons accelerated to the highest energies, close to the Y-point, do not escape as they fall back onto the neutron star surface \citep{Cerutti15}. The maximum Lorentz factors of protons are close to the vacuum polar-cap Lorentz factor of protons $\gamma_{\rm pc, p}$ \citep{Philippov18}. For $f_{\rm pp}=0.05$, in the transition regime between force-free and electrosphere, electrons, positrons, and protons tend to reach higher energies due to higher unscreened electric fields. Regarding particles escaping the magnetosphere, as electrons are more confined in the polar flows, the ones escaping have lower energies. Moreover, the current sheet is thin and thus only a small fraction of positrons escape at the highest energies. A significant fraction of the highest energy protons that are not confined in the current sheet escape the magnetosphere. The proton spectrum shows a peak associated with these escaping protons, which was absent in the force-free regime. For $f_{\rm pp}=0.1$, close to the disc-dome configuration, electrons are accelerated to lower energies due to their confinement in the polar flows. The electrons accelerated to the highest energies do not escape as well. Protons are accelerated to higher Lorentz factors than in the force-free regime, close to the vacuum maximum Lorentz factor of protons $\gamma_{0,\rm p}$.

\subsection{Trajectories and acceleration}

\begin{figure}[t]
\centering
\includegraphics[width=0.49\textwidth]{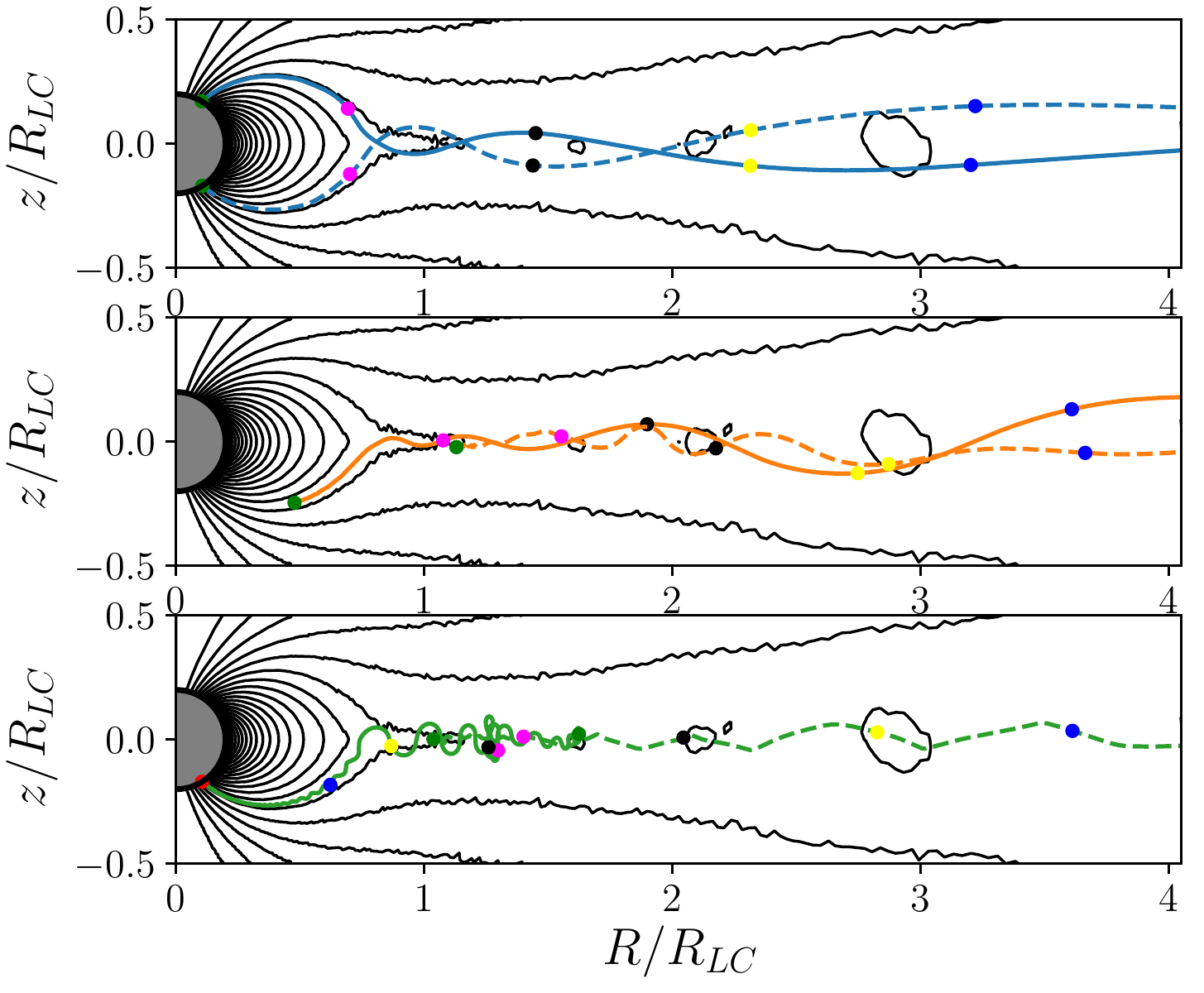}
\caption{Example of proton, positron and electron trajectories (from top to bottom) in the equatorial region for $f_{\rm pp}=0.01$. Markers of different colours link figures~\ref{fig:trajectories} and \ref{fig:part_gamma}.}\label{fig:trajectories}
\end{figure}

\begin{figure}[!t]
\centering
\includegraphics[width=0.49\textwidth]{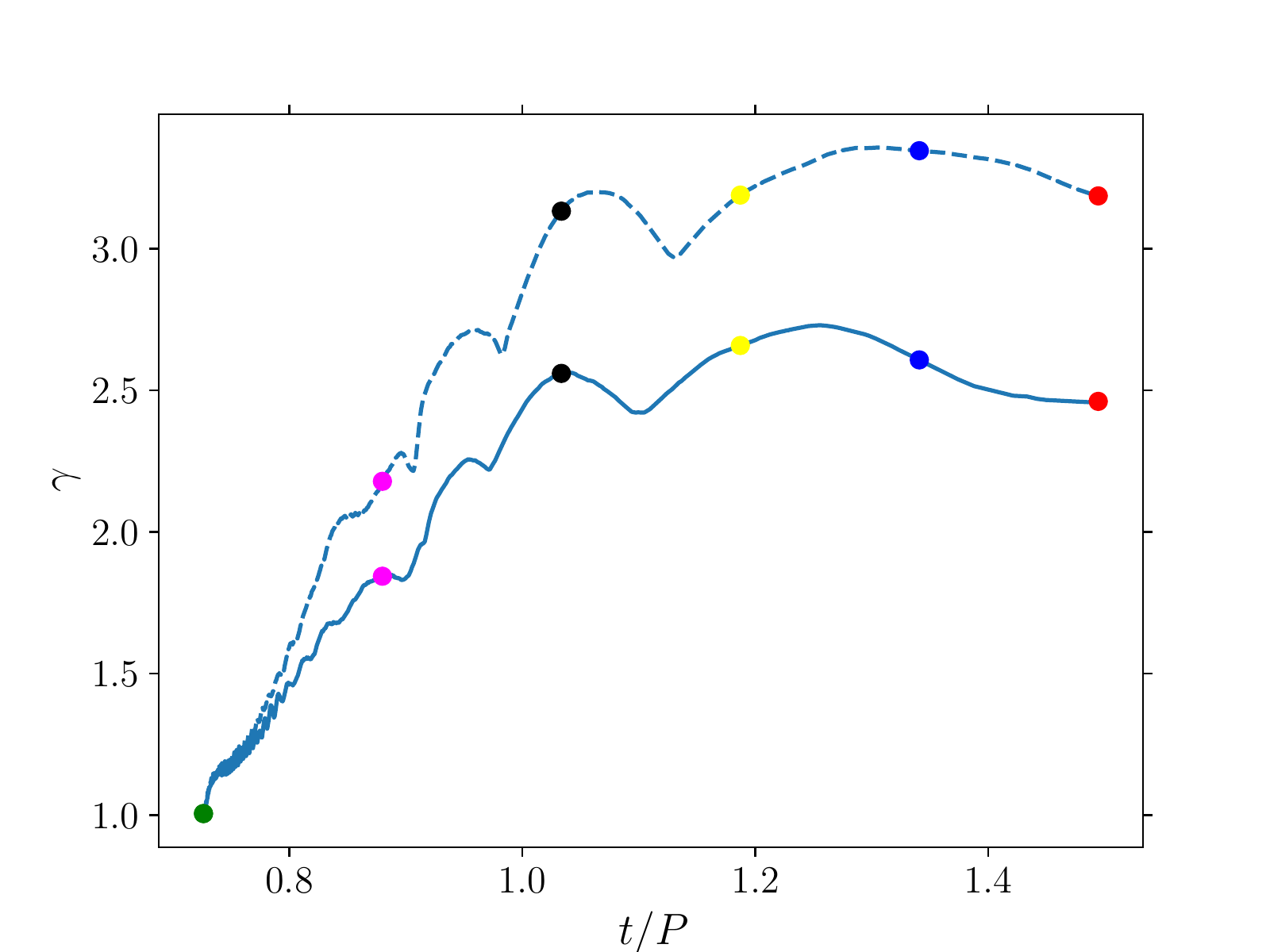}
\includegraphics[width=0.49\textwidth]{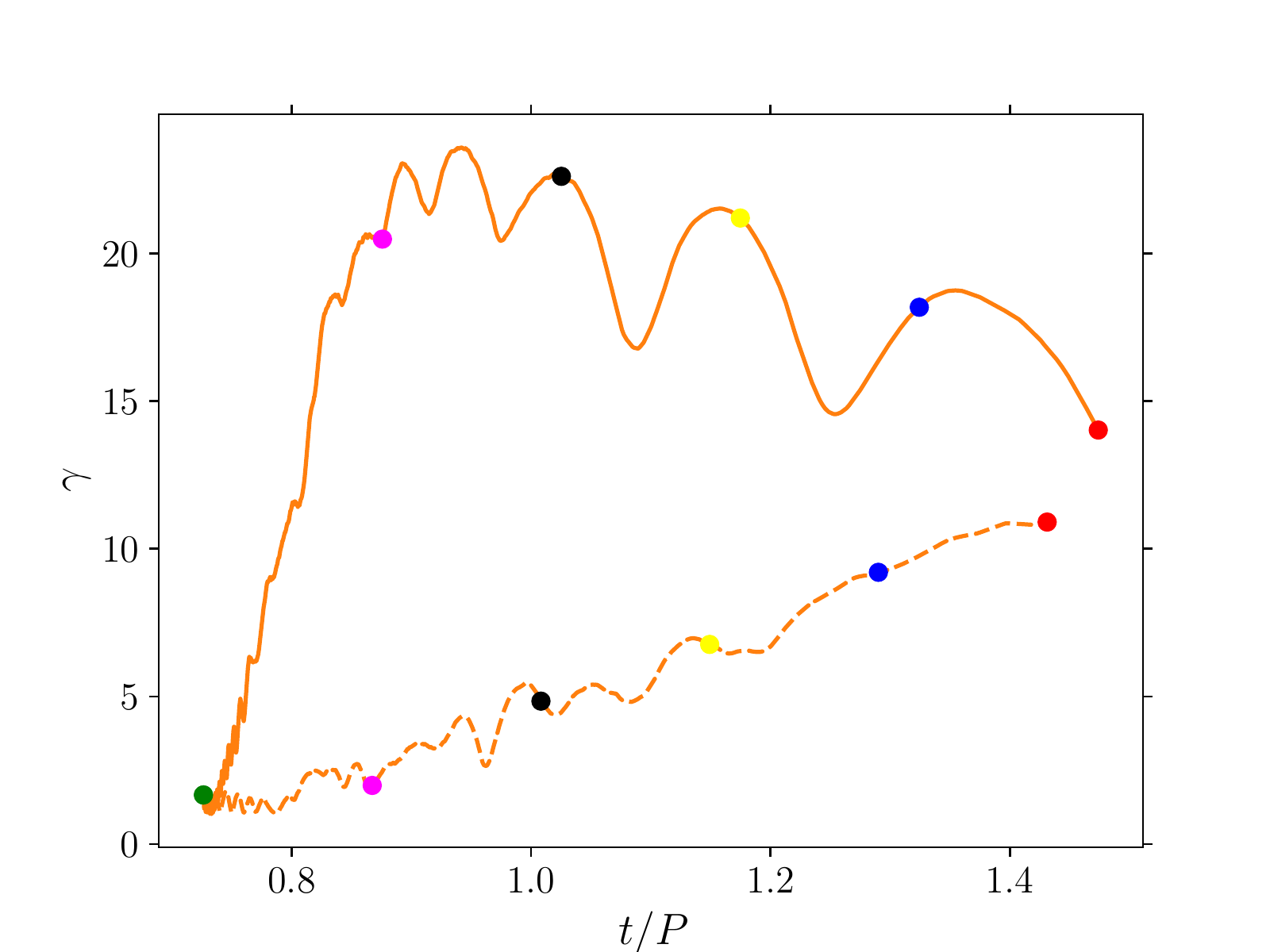}
\includegraphics[width=0.49\textwidth]{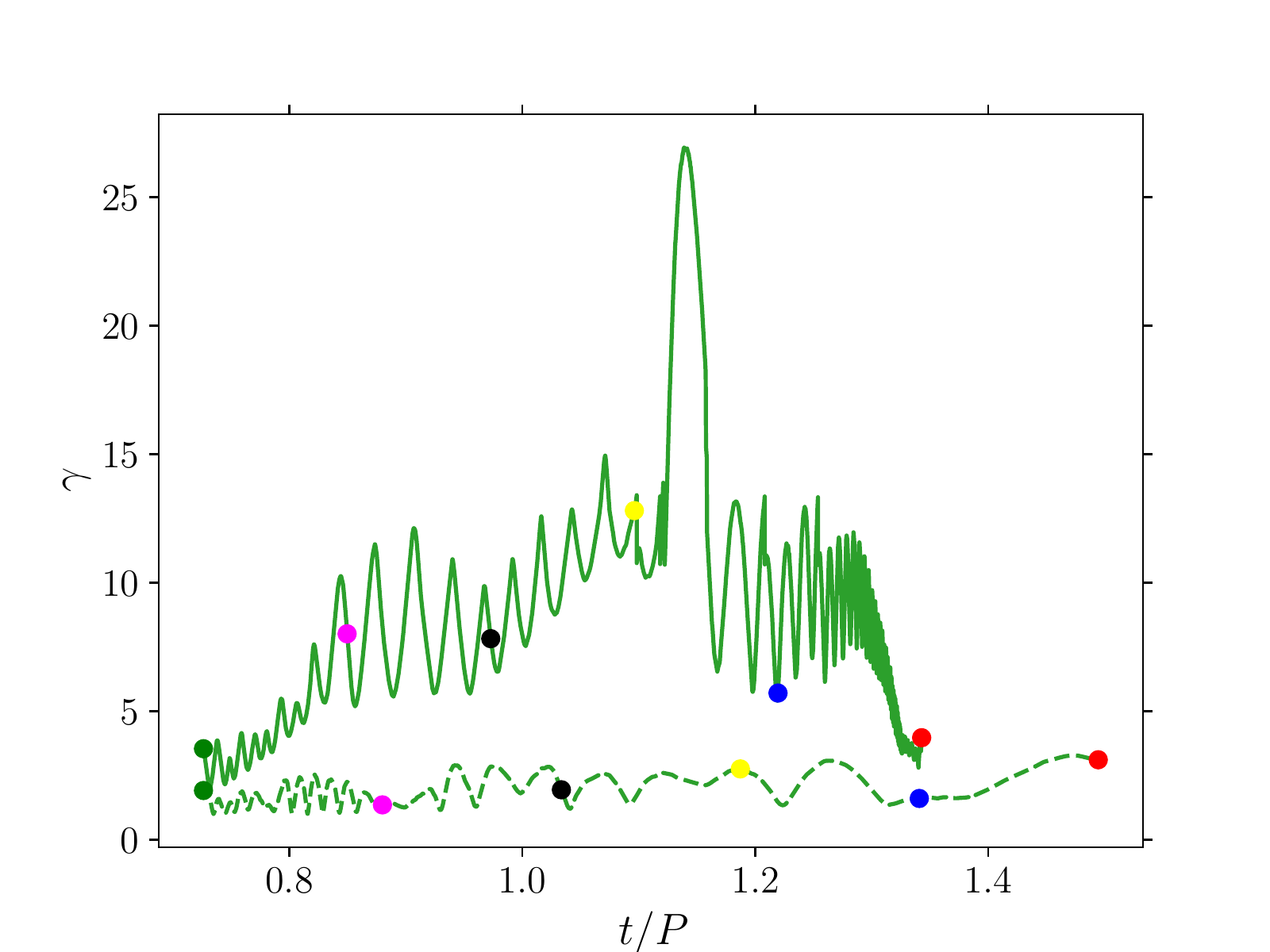}
\caption{Lorentz factors corresponding to the proton, positron, and electron trajectories shown in figure~\ref{fig:trajectories}.}\label{fig:part_gamma}
\end{figure}
In all simulations, a fraction of the injected protons are systematically accelerated and escape the magnetosphere. Most of the protons injected at every time step directly fall back on the neutron star surface. Only protons injected at high latitudes escape, typically at $\theta \sim 0.6-0.7\,{\rm rad}$ (and $\pi-\theta$) for $f_{\rm pp}=0.01$ which coincides with the footpoints of the separatrix current layers, and at $\theta \sim 0.9\,{\rm rad}$ (and $\pi-\theta$) for $f_{\rm pp}=0.1$. In comparison, the polar cap angle is $\theta_{\rm pc} \sim 0.46 \,{\rm rad}$. Protons injected at lower latitudes are trapped in the closed field line region and wrap around the neutron star. Protons injected at the highest latitudes are the ones accelerated to the highest energies and have thus the highest chance of escaping. Protons that escape have quasi-radial trajectories at large distances.

\begin{figure}[!t]
\centering
\includegraphics[width=0.49\textwidth]{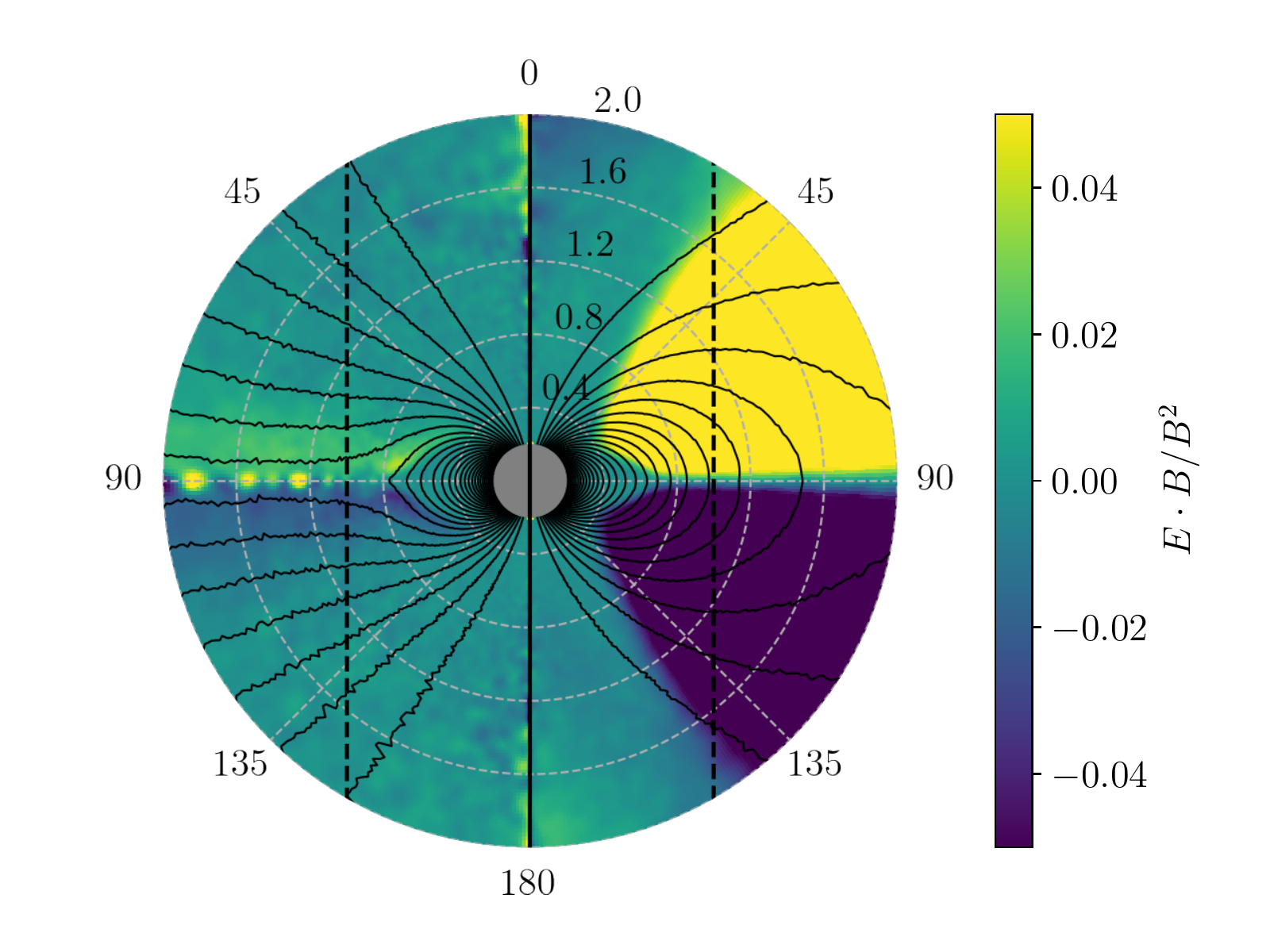}
\caption{Parallel electric field $ \bm{E}\cdot \bm{B}/B^2$ for $r \leq 2 R_{\rm LC}$, for $f_{\rm pp}=0.01$ (left) and $f_{\rm pp}=0.05$ (right).}\label{fig:EdotB}

\vspace{0.5cm}

\centering
\includegraphics[width=0.49\textwidth]{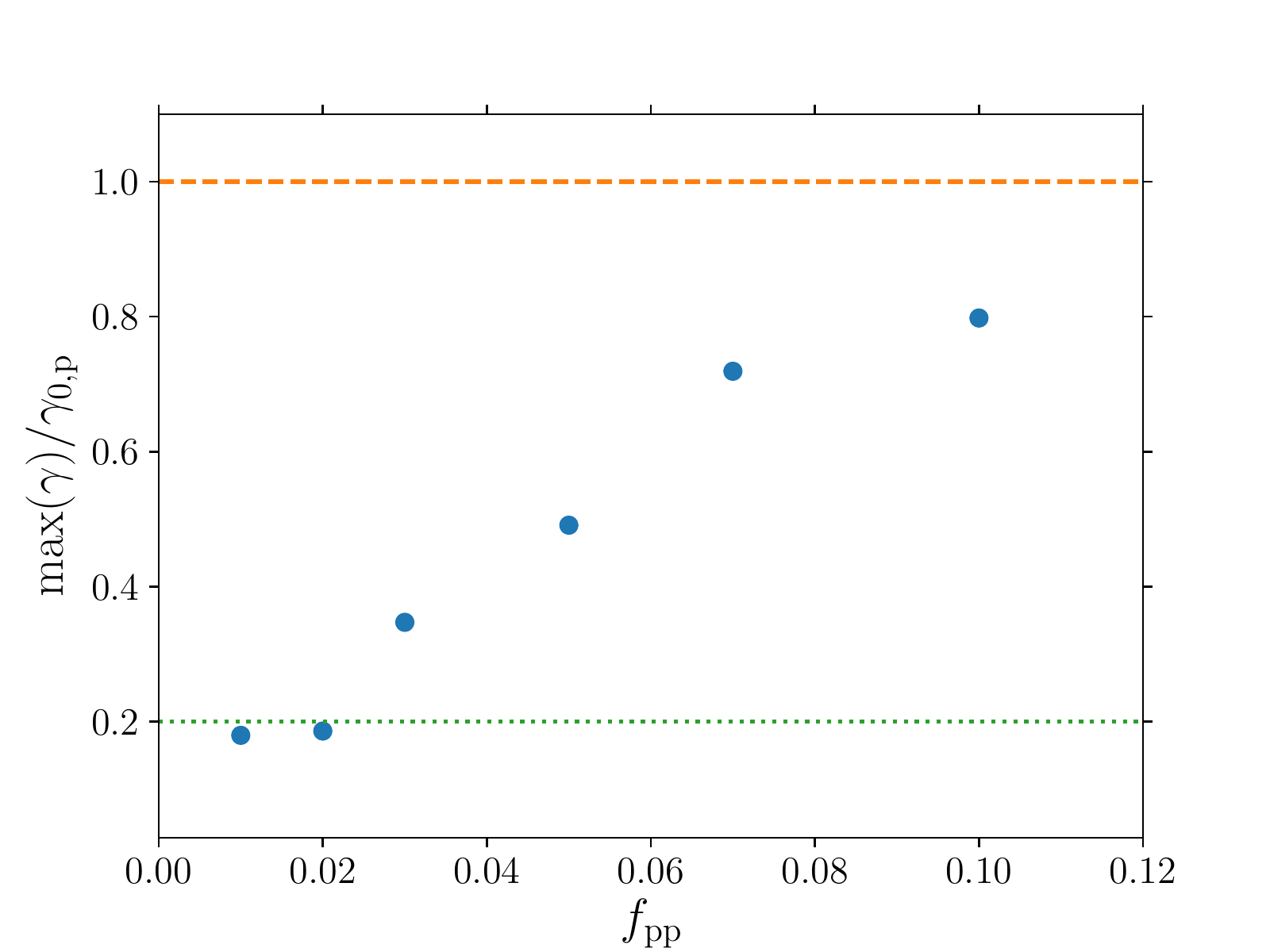}
\includegraphics[width=0.49\textwidth]{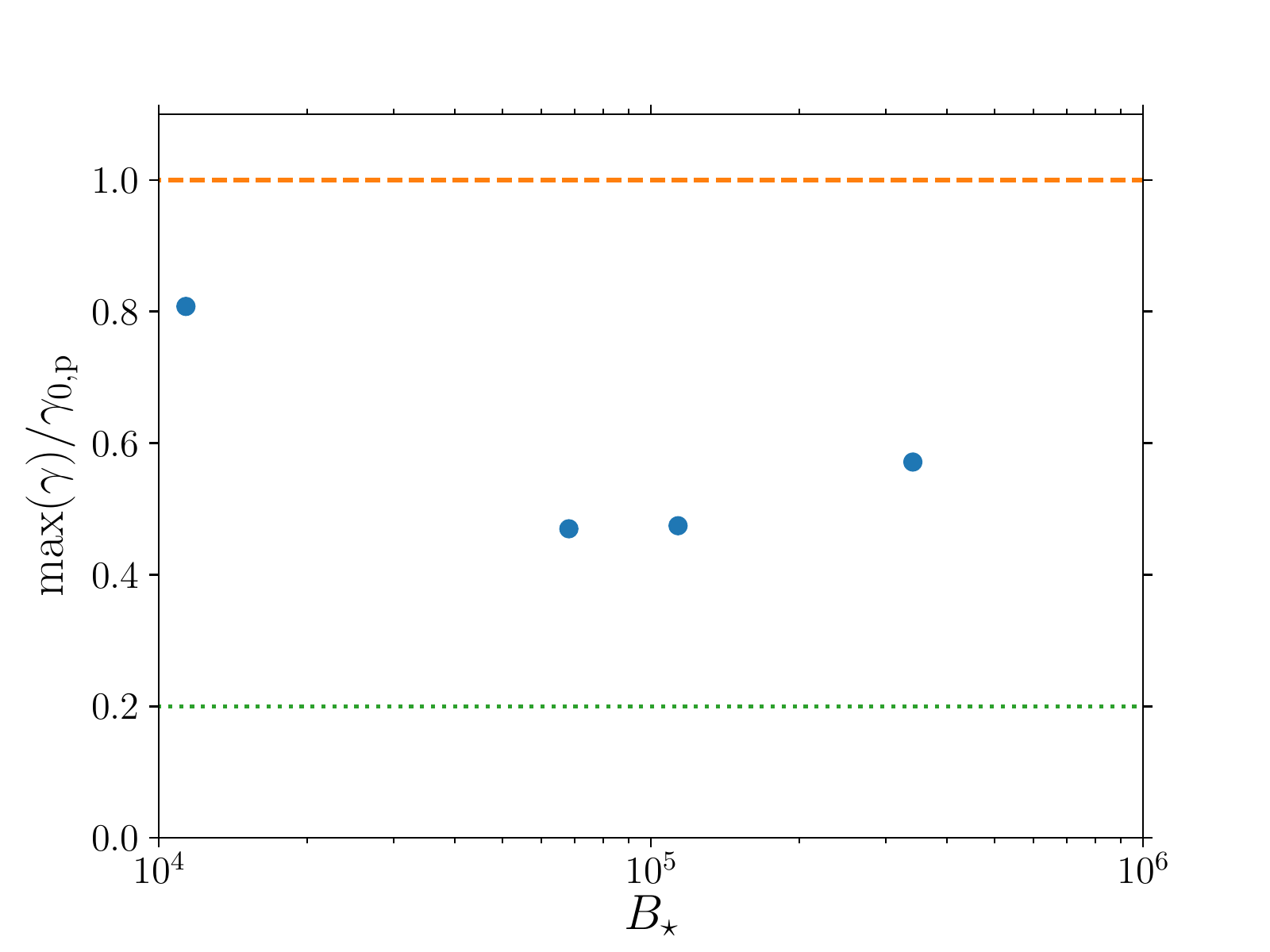}
\caption{Maximum Lorentz factor of escaping protons, normalised by $\gamma_{0,\rm p}$, as a function of $f_{\rm pp}$ for $B_\star = 1.1 \times 10^5\,{\rm G}$ (top) and as a function of $B_\star$ for $f_{\rm pp}=0.05$ (bottom). The dots correspond to simulation results, whereas the lines represent the maximum Lorentz factor of protons experiencing the vacuum potential drop, from pole to equator (orange, dashed) or across the polar cap (green, dotted).}\label{fig:maxgamma}
\vspace{-1cm}
\end{figure}

Trajectories of protons, positrons, and electrons projected in a poloidal plane are illustrated in figure~\ref{fig:trajectories} for $f_{\rm pp}=0.01$. We select two particles for each species, which propagate in the equatorial region where particles are accelerated at the highest energies. Considering individual trajectories of protons and positrons $f_{\rm pp}=0.01$, we note that we retrieve the Lorentz factors by calculating $\gamma \simeq \left[1+ (\int {\rm d}t (e E_\parallel / m c)^2\right]^{0.5}$, where $ E_\parallel = \bm{E} \cdot \bm{v}/v$, with $\bm{v}$ being the particle velocity. As expected, particles are accelerated by the electric field component parallel to their trajectory. We see that protons and positrons are efficiently accelerated  along the separatrices, below the light cylinder radius. Protons can be slightly accelerated at larger distances when they cross the current sheet and experience unscreened parallel electric fields. Their Lorentz factors reach local minima at the crossing points with the current sheet. The positrons are more efficiently confined and thus accelerated in the current sheet, where they acquire a larger fraction of their final Lorentz factors. However, due to the development of instabilities (such as kink and tearing instabilities) and magnetic islands in the current sheet (see figure~\ref{fig:mag_dens}), a wide variety of trajectories can be observed for positrons and electrons. For instance we show the trajectory of one electron that escapes in the equatorial region despite its negative charge (dashed green line). Protons that are less sensitive to the structure of the current sheet display more similar trajectories.

Protons gain a large fraction of their final energy within the separatrix current sheets inside the light-cylinder radius. For a high pair production with $f_{\rm pp}=0.01$, most of the escaping protons gain $75\%$ of their maximum Lorentz factor below $R_{\rm LC}$. For a low pair production with $f_{\rm pp}=0.1$, protons gain $75\%$ of their maximum Lorentz factor below $0.6 R_{\rm LC}$. The fate of positrons is different, as they can be confined in the current sheet and thus accelerated at larger distances, with a significant contribution from magnetic reconnection, which is consistent with previous studies \citep[e.g.][]{Cerutti15}. To more accurately interpret this important result, it is useful to look at the parallel component of the electric field in both magnetospheric regimes, $\mathbf{E}\cdot\mathbf{B}$, as shown in figure~\ref{fig:EdotB}. In the force-free-like solution, $\mathbf{E}\cdot\mathbf{B}\approx 0$ almost everywhere as it should except within the separatrix and equatorial current sheets. Protons fully experience the separatrix electric gap from their injection at the star surface up to the light cylinder where they flow perfectly along the field lines. Beyond the Y-point, they do not experience significant additional acceleration because their trajectories present large oscillations and therefore do not probe the scale of reconnection electric field in the equatorial layer set by the pairs. In contrast, positrons are mostly created outside the light cylinder and are well confined within the equatorial current sheet where they are accelerated by reconnection. In the electrosphere-like configuration, large vacuum gaps fill the magnetosphere outside of the electronic dome and proton disc. As a result, the few protons leaving the dome are quickly accelerated and experience the full vacuum potential.

The maximum Lorentz factor of escaping protons as a function of pair production efficiency is shown in figure~\ref{fig:maxgamma} (top panel). We see that protons experience a fraction of the full vacuum potential drop (higher than the polar cap potential drop for $f_{\rm pp} > 0.01$). This fraction is small for high production of pairs, increases when the production of pairs is reduced, and saturates at a maximum value for low or no pair production $\sim 0.75 \gamma_{0,\rm p}$. The densities of electrons and positrons in the closed field line region, where protons are mostly accelerated, are high for high production of pairs, and therefore the high plasma multiplicities screen the parallel electric field and prevent protons from experiencing a large fraction of the full vacuum potential drop. For low production of pairs, only protons are present in the equatorial plane and can experience a large fraction of the full vacuum potential drop. We note that the magnetic field dependence of the proton maximum Lorentz factor $\gamma_{0, \rm p}$ seems to be well reproduced by the simulations.

In the case of escaping positrons, their maximum Lorentz factor also increases with decreasing pair production (an increasing $f_{\rm pp}$), between $f_{\rm pp} = 0.01$ and $f_{\rm pp} = 0.04$. For lower pair productions $f_{\rm pp} \geq 0.05$, the number of positrons produced strongly decreases, the current sheet does not form, and thus the maximum Lorentz factor of escaping positrons drops.

\subsection{Proton maximum energy in real pulsars}\label{sec:Epmax}

The estimates of the proton Lorentz factors cannot be directly related with realistic cases as the magnetic field, neutron star radius, and mass ratio are downscaled in our numerical experiments. A rescaling procedure is therefore required. In the formula that we use for extrapolation, several quantities intervene such as the radius of the star, the rotation frequency, and the magnetic field. In the range accessible with our simulations, we have performed several series of tests, varying the magnetic field strength $B_\star$ and radius of the neutron star $R_\star$ as well as the mass ratio $m_{\rm r}$ in several sets of simulations in order to check the impact of these parameters on the maximum energy and luminosity. These tests validate the dependencies that intervene in our extrapolation. For instance, as illustrated in the bottom panel of figure~\ref{fig:maxgamma}, the maximum Lorentz factor of protons $\max(\gamma)$ appears to be a nearly constant fraction of $\gamma_{\rm 0,p}$, which suggests that $\max(\gamma)$ is proportional to $B_\star$. We note that the maximum Lorentz factor obtained for the lowest magnetic field is a higher fraction of  $\gamma_{\rm 0,p}$, which is certainly due to the low maximum Lorentz factor and the confusion with thermal protons, as $\gamma_{\rm 0,p} \simeq 3.5$ for $B_\star \sim 10^4\,{\rm G}$. 

Despite these tests, we caution that this rescaling procedure is a delicate process owing to the large difference between numerical and realistic scales. The quantities that we derive should therefore be considered with care. We assume that a constant fraction of the full vacuum potential drop can be channelled into proton acceleration. In our simulations, we obtain maximum Lorentz factors of between $15$ and $75\%$ of $\gamma_{0, \rm p}$, from a high to a low pair production, respectively. As $\gamma_{0, \rm p} = 3.3 \times 10^7\; m^{-1}_{{\rm r},1836} B_{\star,9} R_{\star,6}^2 P_{-3}^{-1}$, we see that protons can be accelerated up to $E_{\rm p} \simeq 5 \times 10^{15} \, {\rm eV} \; B_{\star,9} R_{\star,6}^2 P_{-3}^{-1}$ for a high pair production and up to $E_{\rm p} \simeq 2 \times 10^{16} \, {\rm eV} \; B_{\star,9} R_{\star,6}^2 P_{-3}^{-1}$ for a low pair production. These estimates have been derived for typical properties of millisecond pulsars, with $B_{\star}=10^{9}\,{\rm G}$ and $P=10^{-3}\,{\rm s}$, and a correct electron to proton mass ratio. Thus millisecond pulsars could produce cosmic rays at PeV energies. This could have interesting observational consequences, such as the production of gamma rays in the Galactic centre region \citep{Guepin18a}. For newborn pulsars with millisecond periods, we obtain $E_{\rm p} \simeq 5 \times 10^{19} \, {\rm eV} \; B_{\star,13} R_{\star,6}^2 P_{-3}^{-1}$ for a high pair production and up to $E_{\rm p} \simeq 2 \times 10^{20} \, {\rm eV} \; B_{\star,13} R_{\star,6}^2 P_{-3}^{-1}$ for a low pair production. Therefore, we show that newborn pulsars with millisecond periods could produce cosmic rays up to ultra-high energies, as proposed in several studies \citep{Blasi00, Fang12, Fang13, Lemoine15, Kotera15}. We caution that the effect of curvature radiation is underestimated in our simulations because of the downscaled magnetic field and radius of the neutron star and the subsequent low Lorentz factors of accelerated particles. As curvature radiation can strongly limit particle acceleration below the light cylinder radius \citep{Arons03}, a realistic treatment could therefore impact the acceleration regions and maximum energies of particles, and should therefore be studied in future work. The cases of normal pulsars and millisecond magnetars are difficult to explore with our simulations due to the large distance between the star and the light cylinder radius, or the high magnetic fields. In particular, extreme magnetic fields could have consequences on pair production processes. These configurations therefore require dedicated studies.

\subsection{Energy dissipation and luminosity}

\begin{figure}[!ht]
\centering

\includegraphics[width=0.49\textwidth]{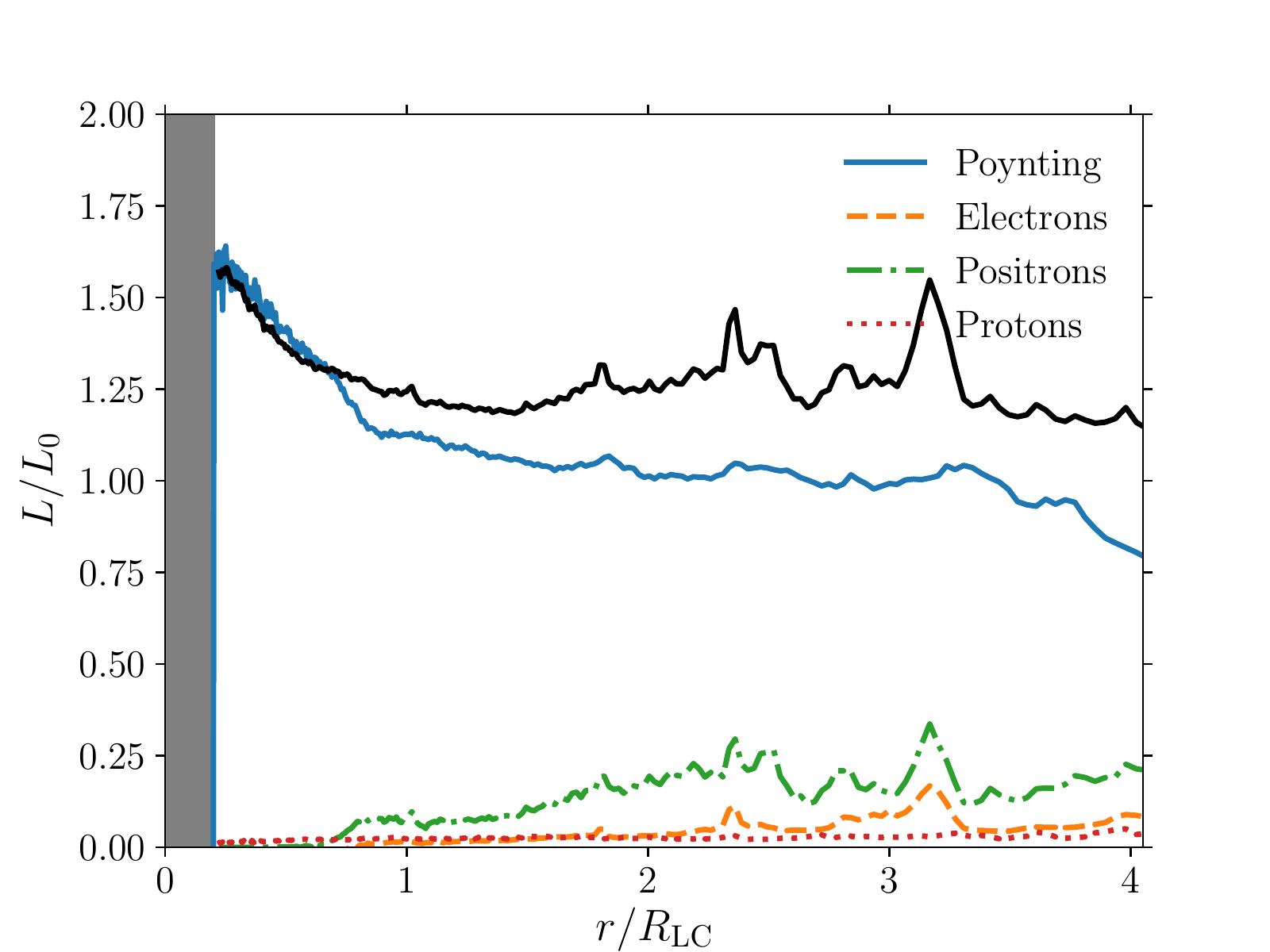}

\vspace{0.5cm}

\includegraphics[width=0.49\textwidth]{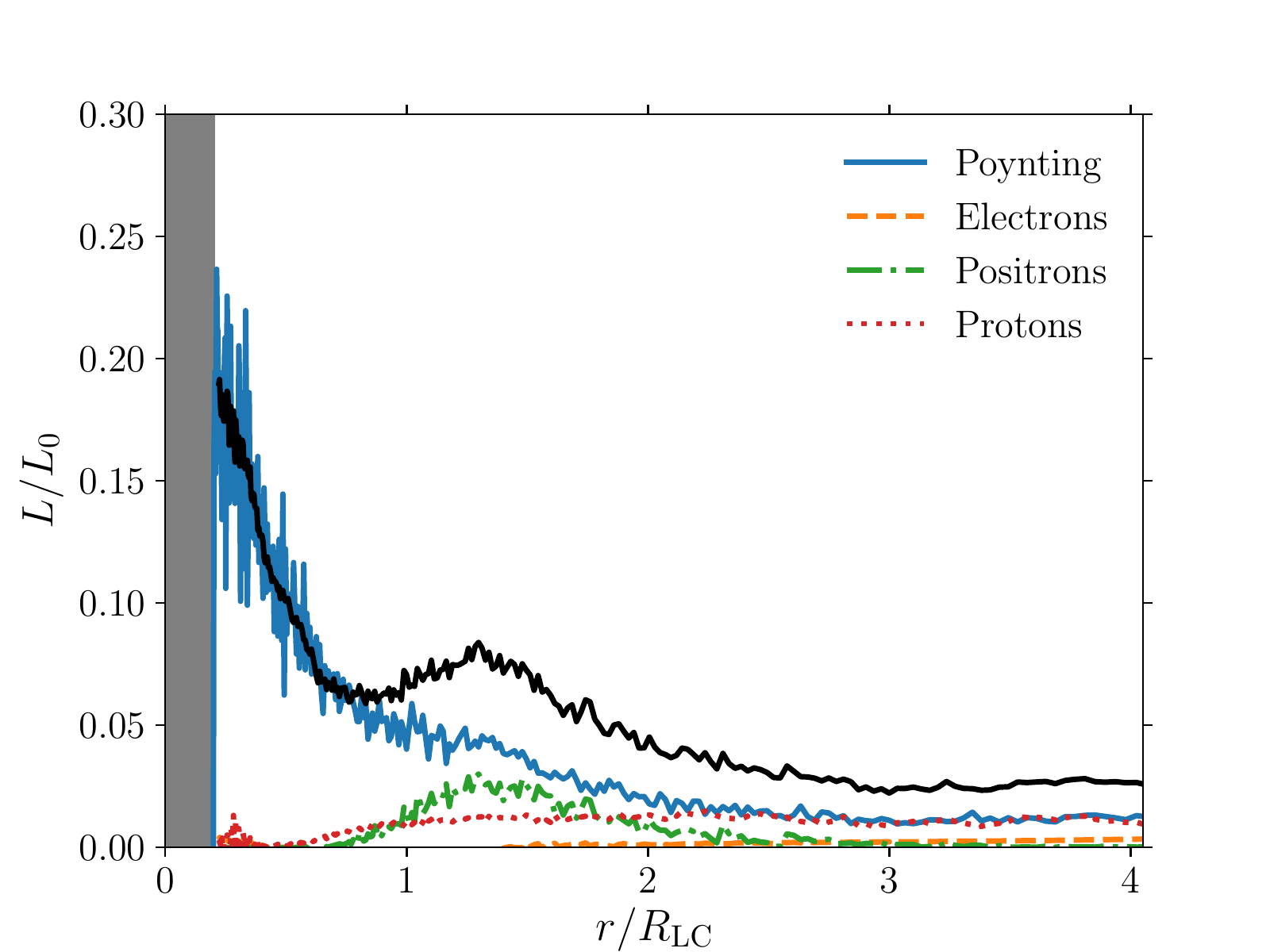}

\vspace{0.5cm}

\includegraphics[width=0.49\textwidth]{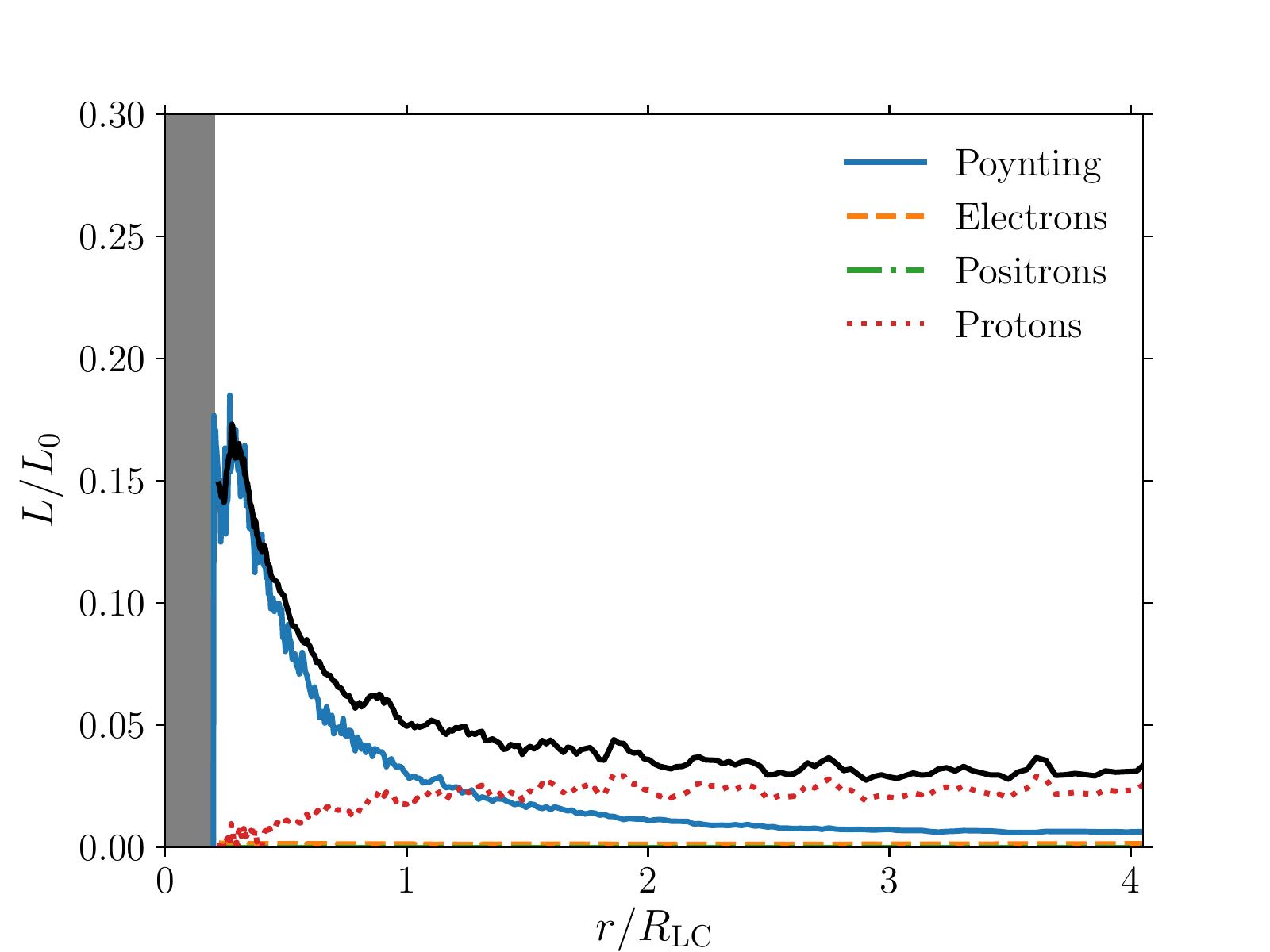}
\caption{Energy dissipation as a function of radius for $f_{\rm pp}=0.01$, $f_{\rm pp}=0.05,$ and $f_{\rm pp}=0.1$ (top to bottom). We show the radial Poynting flux integrated over a sphere of radius $r$ (blue line), the luminosity in electrons (orange dashed line), in positrons (green dot-dashed line), and in protons (red dotted line), and the sum of all these components (black line), normalised by $L_0$. We caution that the vertical scales of the figures are different.}\label{fig:diss}

\vspace{-0.5cm}

\end{figure}

One last important quantity to infer is the total energy dissipated and channelled into particles, which allows us to estimate the proton luminosity. As illustrated in figure~\ref{fig:diss}, the production of pairs has a strong impact on the outgoing Poynting flux; it decreases strongly with a decrease of the yield of pair production. Furthermore, we note that it can be larger than the analytical spin-down power of an aligned pulsar $L_0 = c B_\star^2 R_\star^6 / 4 R_{\rm LC}^4$ \citep[e.g.][]{Contopoulos99,Spitkovsky06} for high pair production, as the Y-point is located below $r=R_{\rm LC}$ and thus a larger fraction of field lines are open. Moreover, it is less than $20\%$ of $L_0$ for low pair productions. Therefore, aligned pulsars with low pair production barely spin-down, as expected for the disc-dome solution \citep{Cerutti15}. 

Energy dissipation is illustrated in figure~\ref{fig:diss}, where we show the radial outgoing Poynting flux and luminosity in electrons, positrons, and protons for $f_{\rm pp}=0.01$, $f_{\rm pp}=0.05$ and $f_{\rm pp}=0.1$ as a function of $r/R_{\rm LC}$. We caution that the scales of the figures are different. These quantities are smoothed over several time-steps and radial bins in order to display  the results clearly. In our simulations, the energy dissipated is self-consistently transferred to particles that are accelerated. A main and irreducible source of magnetic dissipation is via magnetic reconnection which operates in the equatorial current sheet. The separatrices are also a source of dissipation, as they form cavities that allow particle acceleration, with the electric field accelerating particles along the magnetic field lines. The total power shows significant variations with radius, which demonstrates the occurrence of nonstationary phenomena. These irregularities reflect the strong time dependency of reconnection and particle acceleration via the formation of plasmoids (see strong peaks in figure~\ref{fig:diss} for $f_{\rm pp}=0.01$) and kinks in the current sheet, and the related shifts of the Y-point position.

For $f_{\rm pp}=0.01$, the dissipation of radial Poynting flux into particle kinetic energy occurs mostly around and beyond the Y-point, which is located at approximately $r = 0.8 R_{\rm LC}$.  The energy is mostly dissipated into positron kinetic energy. Energy is also dissipated below the Y-point along the gaps where the parallel electric field is not completely screened (see figure~\ref{fig:EdotB}). The fraction of the Poynting flux dissipated into electron and proton kinetic energy decreases with increasing $f_{\rm pp}$. For $f_{\rm pp}=0.1,$ a significant fraction of the Poynting flux is dissipated into proton kinetic energy.

\begin{figure}[t]
\centering
\includegraphics[width=0.49\textwidth]{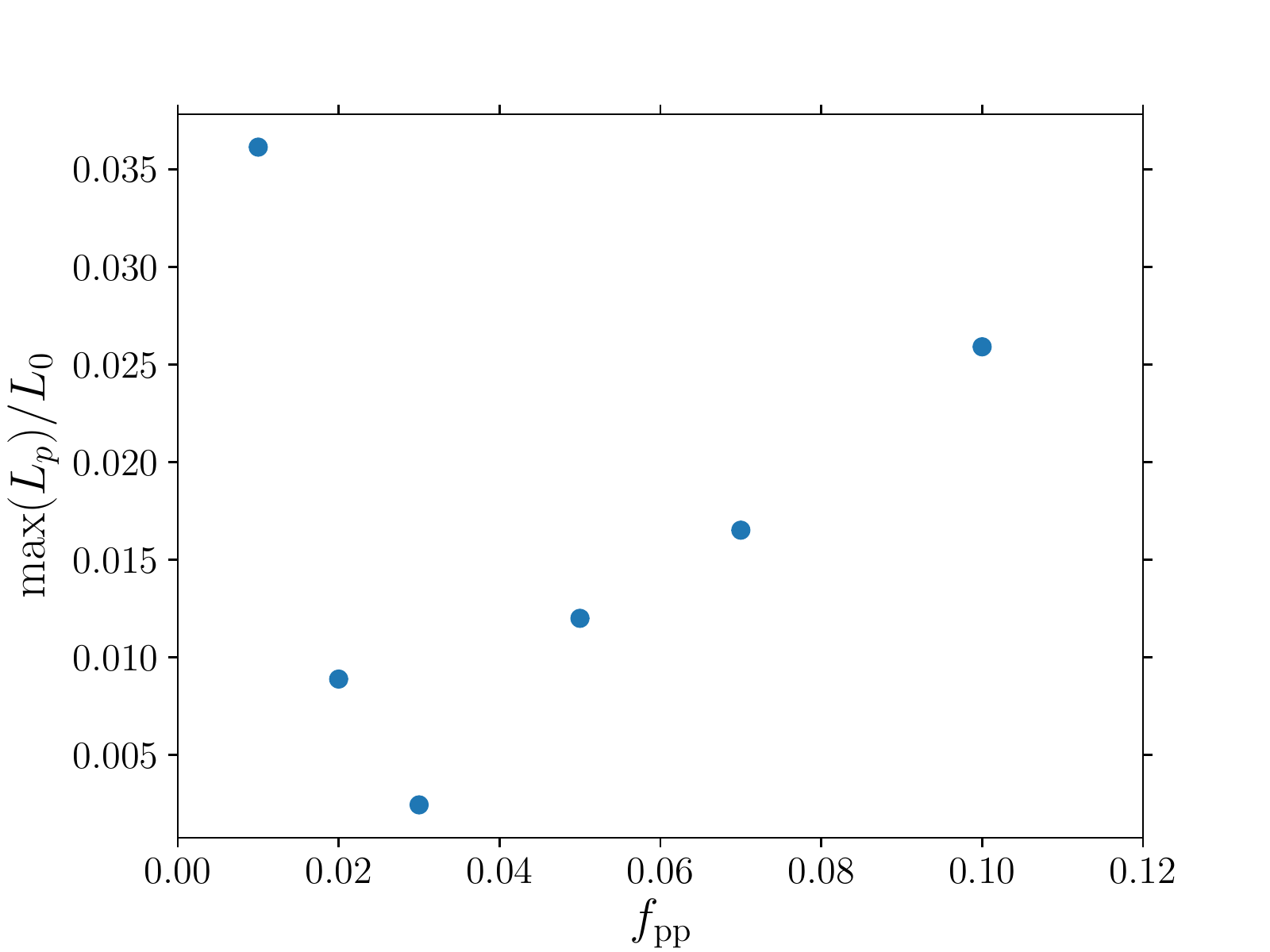}
\caption{Maximum proton luminosity normalised by $L_0$, as a function of $f_{\rm pp}$ for $B_\star = 1.1 \times 10^5\,{\rm G}$.}\label{fig:maxLp}
\end{figure}

These simulations allow us to evaluate the typical proton luminosity. As illustrated in figure~\ref{fig:maxLp}, the maximum proton luminosity is obtained for a high production of pairs. For a decreasing pair production, the proton luminosity first decreases, and then increases again. Given the limited number of simulations that we can perform, it is difficult to determine with certainly the minimum proton luminosity. We obtain the minimum proton luminosity $L_{\rm p} \simeq 2 \times 10^{-3} L_0$ for $f_{\rm pp}=0.03$ and the maximum proton luminosity $L_{\rm p} \simeq 4 \times 10^{-2} L_0$ for $f_{\rm pp}=0.01$. Assuming that we can use these fractions for typical pulsar properties, and considering the value of the spin-down power of an aligned pulsar $L_0 = 1.4 \times 10^{37}\,{\rm erg\,s}^{-1}\; B_{\star,9}^2 R_{\star,6}^6 P_{-3}^{-4}$ for millisecond pulsar properties, we obtain $L_{\rm p} \simeq 3 \times 10^{34}-5 \times 10^{35}\,{\rm erg\,s}^{-1}\; B_{\star,9}^2 R_{\star,6}^6 P_{-3}^{-4}$. For newborn pulsars with millisecond periods, we obtain $L_{\rm p} \simeq 3 \times 10^{42}-5 \times 10^{43}\,{\rm erg\,s}^{-1}\; B_{\star,13}^2 R_{\star,6}^6 P_{-3}^{-4}$.

\section{Discussion and conclusions}\label{Sec:conclusion}

Here, we use 2D PIC simulations to study the impact of pair production on the acceleration of protons in aligned pulsar magnetospheres. These simulations confirm that pulsar magnetospheres are good candidates for the acceleration of protons: regardless of the yield of pair production, protons can be accelerated and escape. Interestingly, due to the mass ratio and large density contrast between protons and pairs, protons do not experience the same trajectories, and thus acceleration, as pairs; they are mostly accelerated below the light cylinder radius within the separatrix current layers but they are not confined in the equatorial current sheet when it exists, whereas pairs are accelerated at their highest energies at the Y-point and beyond in the equatorial current sheet. We note that higher magnetic fields could enhance pair production below the light cylinder radius, as mentioned in \cite{Philippov18}, and screen the parallel electric field that accelerates protons in this region. Thus, protons could be mostly accelerated at larger distances in the current sheet. This effect could be investigated in future studies.

In our numerical experiments, magnetic energy is dissipated into particle kinetic energy through magnetic reconnection in the current sheet and acceleration by the unscreened electric field along the separatrices. Similar dissipation has been observed in different studies with different codes, even using different methods such as resistive force-free MHD \citep[see, e.g.,][]{Parfrey12}. Gamma-ray observations give us a lower limit of the dissipation rate which is given by the GeV power output compared with the pulsar spindown. This rate is comprised between $1$ and $10\%$ and probably even higher in some cases. To this, a bolometric correction factor would be required as well as the efficiency factor of the accelerated particles. The dissipation reported here and in other studies is at least compatible with observed lower limits. For a high production of pairs, about $2\%$ of the theoretical pulsar spin-down power of an aligned pulsar $L_0$ is channelled into protons, which is of the same order as the fraction required to fit the ultra-high-energy cosmic-ray (UHECR) spectrum \citep{Fang13a}. In this case, the maximum particle Lorentz is limited by the potential drop across the pulsar polar cap \citep{Cerutti15,Philippov18}. For a low production of pairs, less than $0.05\%$ of $L_0$ is channelled into protons. However, despite these low luminosities, it appears also that protons are accelerated to higher energies for a low production of pairs. The maximum Lorentz factor approaches and is limited by the theoretical maximum given by the total vacuum potential drop. 

Comparisons between our simulation results and analytical estimates of the maximum Lorentz factor and spin-down luminosity allow us to estimate the maximum energy and luminosity of accelerated protons for realistic parameters. We caution that these estimates should be considered with care owing to the difference between numerical and realistic scales. For typical properties of millisecond pulsars, protons could be accelerated at $E_{\rm p} \simeq 1\,{\rm PeV}$ with luminosities of $L_{\rm p} \simeq  5 \times 10^{35}\,{\rm erg\,s}^{-1}$, which might have interesting observational consequences for the production of gamma rays \citep[e.g.][]{Guepin18a}. Moreover, for typical properties of newborn pulsars with millisecond periods, protons could be accelerated to ultra-high energies, up to $E_{\rm p} \simeq 10\,{\rm EeV}$ with luminosities $L_{\rm p} \simeq  5 \times 10^{43}\,{\rm erg\,s}^{-1}$. 

Typically, a few percent of the spin-down power of the total population of newborn millisecond pulsars is required to reproduce the observed UHECR spectrum \citep[e.g.][]{Fang13}, which seems achievable given our results and therefore supports the idea that newborn pulsars with millisecond periods are good candidate sources for the production of UHECRs. We note that heavy nuclei are required to explain the UHECR spectrum at the highest energies. The extraction of heavy ions from the neutron star surface, as already discussed in \cite{Chen74} and \cite{Kotera15} for instance, as well as their propagation and interaction in the magnetosphere could be explored in a subsequent study. As the neutron star crust is mostly composed of Fe or Ni elements, heavy elements could be naturally extracted from these objects. Extracting ions from the crust could be allowed by the high induced electric fields. Particles bombarding the neutron star surface could also help the extraction.  Moreover, as mentioned in section~\ref{sec:Epmax}, we caution that the energy losses of particles that are due to curvature radiation are not fully taken into account in our simulations. As calculated in \cite{Arons03}, as a consequence of curvature radiation, the energy of accelerated protons below the light-cylinder radius is limited to $E_{\rm p} \simeq 10^{16.5}\,{\rm eV}$ for pulsars with millisecond periods. This effect should be tested in future work in order to assess its impact on numerical simulations and whether or not protons and heavy nuclei can be accelerated at larger distances, in the current sheet or in the wind.

We note that this work is restrictive as only a small fraction of the pulsar wind is included in our simulations, and we do not account for energy losses or re-acceleration of protons at larger distances, for instance at a shock front. Moreover, we consider the case of an aligned pulsar, and the structure of the magnetosphere should be modified in the misaligned case \citep{Spitkovsky06,Petri16}. However, the structure of the magnetosphere below the light cylinder radius, where most of the acceleration of protons seems to take place, should be similar for the misaligned configuration. We highlight that reconnection taking place in the striped wind or Fermi-type acceleration taking place at the termination shock between the pulsar wind and its nebula \citep[e.g.][]{Lemoine15} may further increase the proton maximum energy in the outer regions.

In addition to studying the impact of the pair production strength on proton acceleration in pulsar magnetospheres, we performed several additional tests in order to assess the impact of the size of the simulation domain, the resolution, and the particle injection rate. First, simulations performed by completely shutting down the pair production process lead to the same steady state as the simulations with a low pair production (i.e. $f_{\rm pp}>0.15$). The structure of magnetospheres with a low production of pairs appears to be more affected by simulation parameters: the size of the simulation domain influences the extent in latitude of the polar flows of electrons, and the resolution influences the density of the equatorial flow of protons. Moreover, the injection rate influences these two characteristics. 
In particular, we noticed that simulations performed with a lower resolution show a larger number of protons escaping in the equatorial flow. This difference could result from numerical effects, or the longer times required to reach the steady state for higher resolutions, and is currently under study. However, despite small morphological differences, the general structure of the magnetosphere is similar and our main conclusions remain unchanged. In particular, the maximum energy of particles and the acceleration regions are not affected by these minor structural changes.

Further work will be required to better characterise the escape of protons by a detailed modelling of their trajectories. Moreover, the link between the simulated amount of pair production and pair multiplicities in realistic environments should be explored. For this purpose, a self-consistent modelling of pair production, but also of other types of interactions, will be required. The present study focuses on the aligned magnetosphere. The anti-aligned and inclined cases could be studied in future work. The anti-aligned configuration would reverse the charge densities at the surface of the neutron star, allowing for proton extraction at the poles and electron extraction in the equatorial region. This configuration could lead to interesting magnetospheric configurations because of the mass separation between protons and electrons. Given the fate of the electrons extracted from the poles in the aligned case, the acceleration of protons in the anti-aligned case could be suppressed. This interesting question should be investigated in a dedicated study.

\section*{Acknowledgement}
The authors thank the anonymous referee for useful feedback on this work. We thank A. Spitkovsky, A. Philippov, F. Mottez and J. P\'etri as well as the organizers and participants to the ``Entretiens pulsars'' for useful discussions. This work is supported by the APACHE grant (ANR-16-CE31-0001) of the French Agence Nationale de la Recherche as well as CNES. CG is supported by a fellowship from the CFM Foundation for Research and by the Labex ILP (reference ANR-10-LABX-63, ANR-11-IDEX-0004-02). This work was granted access to the HPC resources of CINES on Occigen under the allocations A0030407669 and A0050407669 made by GENCI. This work has made use of the Horizon Cluster hosted by Institut d’Astrophysique de Paris. We thank Stephane Rouberol for running smoothly this cluster for us.

\bibliographystyle{aa}
\bibliography{PulsarMag}

\appendix

\end{document}